\documentclass[%
 reprint,
 amsmath,amssymb,
 aps,
]{revtex4-2}

\usepackage{graphicx}% Include figure files
\usepackage{dcolumn}% Align table columns on decimal point
\usepackage{bm}% bold math
\usepackage{balance}
\def\Xint#1{\mathchoice
   {\XXint\displaystyle\textstyle{#1}}%
   {\XXint\textstyle\scriptstyle{#1}}%
   {\XXint\scriptstyle\scriptscriptstyle{#1}}%
   {\XXint\scriptscriptstyle\scriptscriptstyle{#1}}%
   \!\int}
\def\XXint#1#2#3{{\setbox0=\hbox{$#1{#2#3}{\int}$}
     \vcenter{\hbox{$#2#3$}}\kern-.5\wd0}}

\def\dashint{\Xint-}

\begin{document}

\preprint{APS/123-QED}

\title{A Generalization of the Kramers-Kronig Relations for Linear Time-Varying Media}% Force line breaks with \\

\author{Diego M. Sol\'{i}s}
\author{Nader Engheta}%
\affiliation{%
 Department of Electrical and Systems Engineering, University of Pennsylvania, Philadelphia, Pennsylvania, USA
}%

\date{\today}% It is always \today, today,

\begin{abstract}
We explore the mathematical theory to rigorously describe the response of media with linear time-varying, generally dispersive, electromagnetic constitutive parameters. We show that, even when the temporal inhomogeneity takes place on a time scale comparable---or shorter---than the driving fields' time period, one can still define a physically meaningful time-varying dispersion. Accordingly, a generalized set of Kramers-Kronig relations is investigated to link the real and imaginary parts of the time-varying frequency-dispersive spectra characterizing the medium's constitutive response. Among others, we study the case of a Lorentzian dielectric response with time-varying volumetric density of polarizable atoms and present the varying circuital equivalents of the governing differential equation, which in turn allow us to use the notion of generalized time-varying impedances/admittances of a time-dependent resistor, inductor and capacitor.
\end{abstract}

\maketitle

%\tableofcontents

\section{\label{sec:level1}Introduction}

The field of ``dynamic'' (i.e., time-variant) metamaterials has recently emerged within the metamaterial community and is rapidly expanding as the next generation of metamaterials. Their predecessors, ``static'' (i.e., time-invariant) metamaterials \cite{engheta2006metamaterials}, are regarded as artificial materials engineered through the (deeply) subwavelength---both in size and periodicity---space-variation of some of their physical properties, such as permittivity or permeability, which opens up fascinating possibilities in harnessing light in ways that were unimaginable years ago, although photonic crystals \cite{joannopoulos1997photonic} (with spatial features electrically larger than in metamaterials) already allowed for other interesting phenomena like bandgaps \cite{1991PhRvL..67.2295Y,1996Natur.383..699K} and slow light \cite{PhysRevLett.87.253902}. Time-invariant metamaterials gave rise, in the first decade of the 21st century, to new paradigms in the way the electromagnetic waves interact with matter, including left-handedness \cite{Veselago_1968,Shelby77}, cloaking \cite{PhysRevE.72.016623,Leonhardt1777,Schurig977}, epsilon-near-zero media \cite{PhysRevLett.97.157403,PhysRevLett.100.033903,Liberal2017}, and magnetless nonreciprocity \cite{doi:10.1063/1.3615688}, to name a few. Their two-dimensional equivalents, "static" metasurfaces, have also gained a lot of momentum in this present decade, given their ability to tailor the amplitude, phase and polarization of waves, yet without the bulkiness or loss-related limitations of their three-dimensional counterparts. A plethora of metasurface-supported exotic effects and applications have been reported, e.g., strong nonlinear responses \cite{Lee2014}, dramatic enhancement of the local density of states via hyperbolic dispersion \cite{PhysRevLett.114.233901}, photonic topological states \cite{Gorlach2018}, all-optical real-time signal processing \cite{Silva160}, angular filtering \cite{Shen1499}, and photonic quantum vortices \cite{doi:10.1021/nl500658n}.

Dynamic metamaterials (and metasurfaces) add yet another degree of freedom and controllability by inducing, with some external source of energy, a temporal change in some of the materials' properties and are therefore spatio-temporally variant. Despite the embryonic stage of this research field, there are already several examples of possible applications enabled by such time-varying materials, ranging from more efficient frequency mixers \cite{PhysRevB.97.115131} and matching-networks \cite{PhysRevLett.121.204301} to spatio-temporal-based nonreciprocity for magnetic-free optical isolators/circulators \cite{Yu2009,doi:10.1021/ph400058y}, angularly-selective nonreciprocal transmission \cite{PhysRevB.92.100304}, time-reversal mirrors \cite{Bacot_2016}, and antireflection coatings \cite{Pacheco-Pena:20}. Moreover, seeing as time-invariant spatially inhomogeneous metastructures have proved to perform mathematical operations \cite{Silva160,estakhri2019inverse}, the idea of adding time variation to expand the range of applicability of these metastructures to, e.g., the linear compansion of a pulse \cite{8889149} is especially promising.

One possible way of achieving time variation is by temporally-modulating (electro-optically, for instance) the dielectric function of a medium. In \cite{PhysRevLett.32.1101}, a high-power electromagnetic pulse was reported to ionize a plasma, creating a nonstationary interface. This rapid change in the dielectric permittivity produces a ``time interface'' or step transient \cite{1124533,Xiao:14,8858032} which, from temporal continuity considerations for both the electric displacement and the magnetic induction fields, is seen to produce frequency-shifted forward and backward waves described by the time-equivalent of the Fresnel coefficients. Time-periodic inhomogeneities in the dielectric function were later addressed in the context of wave propagation in an unbounded medium \cite{1138637}, a half-space \cite{1139931,86907}, or a spacetime-periodic medium \cite{5252018}. In this regard, it is well-known that the application of the Bloch-Floquet theorem to time-periodic media yields a frequency-periodic band-structured dispersion relation exhibiting forbidden wavevector gaps \cite{PhysRevA.79.053821}, dual of the bandgaps found in space-periodic media. These scenarios contemplate, however, only nondispersive electric susceptibilities.

In this manuscript, we focus on the local (i.e., wave propagation is not the object of study here) dynamics of media with a susceptibility that is time-variant and dispersive in general, and we study the transient behavior of the polarization that arises under such conditions. More specifically, we tackle the problem by adopting the methodology of linear systems theory in order to develop rigorous mathematical tools enabling us to investigate the time-varying impulse response of such temporal media, allowing us to generalize the Kramers-Kronig relations \cite{deL.Kronig:26,kramers1927diffusion} for non-instantaneous time-varying media.

\section{Theoretical Analysis}
\newcommand{\comment}[1]{}
There is a well-developed body of knowledge describing linear time-variant (LTV) channels in the signal processing community, inasmuch as mobile communications rely on multipath fading channels modeled as time-variant linear filters \cite{hlawatsch2011wireless,proakis2008digital} (in this regard, we should mention that extensive research work has also been done on time-varying circuits, for example in the context of control engineering---both from the perspective of functional analysis \cite{1701220,spaulding1964passive,1454337} and state-space theory \cite{zadeh1963linear}---or the model reduction of the time-varying equivalents that result from linearizing nonlinear circuits \cite{1103333,799678,1174093}). We will therefore borrow the mathematical apparatus describing multipath propagation and start by writing the response of an LTV system to an arbitrary input $x(t)$ as the following Fredholm integral
\begin{equation}
y(t)=\int_{-\infty}^{\infty}h(t,\tau)x(\tau)d\tau,\label{eq:1}
\end{equation}
in which case it is straightforward to see that $h(t,\tau)$ is defined as the system response at time $t$ to an impulse applied at time $\tau$ (note that, throughout this text, we intentionally leave the upper integration limit of the superposition integrals go to $+\infty$, i.e., we choose to define these integrals in a general form by making no \textit{a priori} assumptions on the causality of their kernels). Indeed, inserting $x(t)$=$\delta(t-\tau)$ into the previous integral and changing the integration variable from $\tau$ to $\tau'$ for ease of notation, one can write
\begin{equation}
y(t)=\int_{-\infty}^{\infty}h(t,\tau')\delta(\tau'-\tau)d\tau'=h(t,\tau).\label{eq:2}
\end{equation}
It is oftentimes more convenient to resort to an alternative formulation as introduced by Kailath \cite{baghdady1961lectures} and use instead $\hat{\tau}$=$t-\tau$. A change of variable in Eq.~(\ref{eq:1}) yields
\begin{equation}
y(t)=\int_{-\infty}^{\infty}h(t,t-\hat{\tau})x(t-\hat{\tau})d\hat{\tau}=\int_{-\infty}^{\infty}c(t,\hat{\tau})x(t-\hat{\tau})d\hat{\tau},\label{eq:3}
\end{equation}
where $c(t,\hat{\tau})$, known as the input delay-spread function \cite{1088793}, is now the response measured at $t$ due to an impulse applied at $t-\hat{\tau}$. Proceeding similarly as before we have, for $x(\hat{\tau}')$=$\delta(\hat{\tau}'-(t-\hat{\tau}))$,
\begin{equation}
\begin{split}
y(t)&=\int_{-\infty}^{\infty}c(t,\hat{\tau}')\delta(t-\hat{\tau}'-(t-\hat{\tau})d\hat{\tau}' \\
&=\int_{-\infty}^{\infty}c(t,\hat{\tau}')\delta(\hat{\tau}-\hat{\tau}')d\hat{\tau}'=c(t,\hat{\tau}).\label{eq:4}
\end{split}
\end{equation}

\comment{
\begin{flalign}
y(t) &= \int_{-\infty}^{\infty}c(t,\hat{\tau})\delta(t-\hat{\tau}'-(t-\hat{\tau})d\hat{\tau}'&&\nonumber\\
            &= \int_{-\infty}^{\infty}c(t,\hat{\tau})\delta(\hat{\tau}-\hat{\tau}')d\hat{\tau}'&&\label{eq:4}
\end{flalign}

\begin{multline}
y(t)=\int_{-\infty}^{\infty}c(t,\hat{\tau})\delta(t-\hat{\tau}'-(t-\hat{\tau})d\hat{\tau}' \\
=\int_{-\infty}^{\infty}c(t,\hat{\tau})\delta(\hat{\tau}-\hat{\tau}')d\hat{\tau}',\label{eq:4}
\end{multline}

\begin{eqnarray}
y(t)=\int_{-\infty}^{\infty}c(t,\hat{\tau})\delta(t-\hat{\tau}'-(t-\hat{\tau})d\hat{\tau}'\nonumber \\ =\int_{-\infty}^{\infty}c(t,\hat{\tau})\delta(\hat{\tau}-\hat{\tau}')d\hat{\tau}'
\label{eq:4}
\end{eqnarray}
}

In essence, $c(t,\hat{\tau})$=$c(t,t-\tau)$ moves the impulse time frame reference from the origin over to $t$, very much in the same way a Green's tensor does in a translationally invariant space domain when we write $\bar{\bar{G}}(\bold{r},\bold{r}')$=$\bar{\bar{G}}(\bold{r}-\bold{r}')$, with $\bold{r}$ and $\bold{r}'$ observation and source positions, respectively. One can immediately see that $h(t,\tau)$=$h(t-\tau)$ and $c(t,\hat{\tau})$=$c(\hat{\tau})$ when there is time invariance. Besides, it is clear that causality implies $h(t,\tau)$=$0$ for $t\!<\!\tau$ or, alternatively, $c(t,\hat{\tau})$=$0$ for $\hat{\tau}\!<\!0$. Additionally, if $c(t,\hat{\tau})$ is separable, i.e., $c(t,\hat{\tau})$=$c_{t}(t)c_{\hat{\tau}}(\hat{\tau})$, Eq.~(\ref{eq:3}) is simplified to
\begin{equation}
y(t)=c_{t}(t)\int_{-\infty}^{\infty}c_{\hat{\tau}}(\hat{\tau})x(t-\hat{\tau})d\hat{\tau}=c_{t}(t)\Big(c_{\hat{\tau}}(t)\underset{t}{*}x(t)\Big).\label{eq:5}
\end{equation}

Let us focus the discussion on time-varying systems that can be characterized as a linear differential equation with time-varying coefficients of the form \cite{kaplan1962operational}
\begin{equation}
a_n(t)\frac{d^ny(t)}{dt^n}+...+a_1(t)\frac{dy(t)}{dt}+a_0(t)y(t)=x(t).\label{eq:2Z1}
\end{equation}
The electric response of a linear dispersive time-varying medium characterized by a Lorentzian resonance, but whose volume density of polarizable atoms $N(t)$ is time-dependent, falls under this category. This is one of the simplest scenarios one can think of, since the relative amplitude of the coefficients in Eq.~(\ref{eq:2Z1}) remains unperturbed, as seen below:  
\begin{equation}
\frac{d^2P(t)}{dt^2}+\gamma\frac{dP(t)}{dt}+\omega_0^2P(t)=\epsilon_0\omega_p^2(t)E(t),\label{eq:2Z2}
\end{equation}
$E$ and $P$ being the electric field and (local) linear polarization, respectively, and $\omega_p(t)\propto\sqrt{N(t)}$ the plasma frequency. This is equivalent to a linear time-invariant (LTI) medium that responds to $\omega_p^2(t)E(t)$ rather than to $E(t)$. If we define $A(t)$=$\omega_p^2(t)$, this translates in the frequency domain to 
\begin{equation}
P(\omega)=\epsilon_0\frac{\frac{1}{2\pi}A(\omega)\underset{\omega}{*}E(\omega)}{\omega_0^2-\omega^2+i\gamma\omega},\label{eq:2Z3}
\end{equation}
where $\underset{\omega}{*}$ denotes the convolution operation with respect to $\omega$ and $e^{i\omega t}$ convention is chosen. One can arrive at a wave equation for $E(z,t)$ of the form
\begin{equation}
\begin{split}
&\left(\frac{\partial ^2}{\partial t^2}+\gamma\frac{\partial}{\partial t}+\omega_0^2\right)\left(\frac{\partial ^2E(z,t)}{\partial z^2} - \frac{1}{c^2}\frac{\partial ^2E(z,t)}{\partial t^2}\right) \\
&=\frac{1}{c^2}\frac{\partial ^2\big(A(t)E(z,t)\big)}{\partial t^2},\label{eq:2Z4}
\end{split}
\end{equation}
which collapses to the wave equation for a time-varying lossless plasma (i.e., Drude-type medium) ($\omega_0$=$0$, $\gamma$=$0$) in \cite{1139657}:
\begin{equation}
\frac{\partial ^2E(z,t)}{\partial z^2} - \frac{1}{c^2}\frac{\partial ^2E(z,t)}{\partial t^2}=\frac{1}{c^2}A(t)E(z,t).\label{eq:2Z5}
\end{equation}
If one wants to express Eq.~(\ref{eq:2Z2}) in terms of the system's input response as in Eq.~(\ref{eq:1}), $P(t)$=$\epsilon_0\int_{-\infty}^{\infty}\chi_h(t,\tau)E(\tau)d\tau$, it suffices to recognize the time-invariance equivalence mentioned earlier: $P(t)$=$\epsilon_0\int_{-\infty}^{\infty}\chi(t-\tau)A(\tau)E(\tau)d\tau$ (incidentally, note that this situation is different than the one depicted in \cite{mirmoosa2020dipole}, which is rather described by $P(t)$=$\epsilon_0A(t)\int_{-\infty}^{\infty}\chi(t-\tau)E(\tau)d\tau$). It thus follows from inspection that $\chi_h(t,\tau)$=$\chi(t-\tau)A(\tau)$, with 
\comment{
$\chi_h^{(1)}(t,\tau)$=$\chi^{(1)}(t-\tau)\omega_p^2(\tau)$
}
\begin{equation}
\chi(t)=\frac{1}{\sqrt{\omega_0^2-(\gamma/2)^2}}e^{-(\gamma/2) t}\text{sin}\left(t\sqrt{\omega_0^2-(\gamma/2)^2}\right)U(t),\label{eq:2Z6}
\end{equation}

\comment{
\begin{equation}
\mathcal{L}(t)=\frac{1}{\sqrt{\omega_0^2-\gamma^2}}e^{-\gamma t}\text{sin}\left(t\sqrt{\omega_0^2-\gamma^2}\right)U(t),\label{eq:2Z6}
\end{equation}
}

\comment{
\begin{equation}
\begin{split}
\chi_h^{(1)}(t,\tau)=\frac{\omega_p^2(\tau)}{\sqrt{\omega_0^2-\gamma^2}}e^{-\gamma(t-\tau)}\text{sin}\left((t-\tau)\sqrt{\omega_0^2-\gamma^2}\right) \\ \cdot U(t-\tau),\label{eq:2Z6}
\end{split}
\end{equation}
}

\comment{
\begin{equation}
\begin{split}
\chi_h^{(1)}(t,\tau)=\frac{e^{-\frac{\gamma}{2}(t-\tau)}}{\sqrt{\omega_0^2-\left(\frac{\gamma}{2}\right)^2}}\text{sin}\left((t-\tau)\sqrt{\omega_0^2-\left(\frac{\gamma}{2}\right)^2}\right)\\\omega_p^2(\tau)U(t-\tau),\label{eq:2Z6}
\end{split}
\end{equation}
}
where we have used the well-known result for a time-invariant Lorentzian medium and $U(t)$ is the step function. It is compelling to point out that the response to an impulse applied at $\tau$ is not a function of how $N(t)$ evolves for $t>\tau$; this is traced back to the relative weights of the coefficients in Eq.~(\ref{eq:2Z1}) being invariant. Perhaps a circuital analogy would be of use here to better understand this behavior: this Lorentzian response can be thought of as the (polarization) charge response to an applied voltage across a time-varying series RLC circuit, such that Eq.~(\ref{eq:2Z2}) is recast to
\begin{equation}
L(t)\frac{d^2P(t)}{dt^2}+\left(R(t)+\frac{dL(t)}{dt}\right)\frac{dP(t)}{dt}+\frac{1}{C(t)}P(t)=E(t),\label{eq:2Z7}
\end{equation}
with $L(t)$=$\frac{1}{\epsilon_0\omega_p^2(t)}$, $R(t)$=$\gamma L(t)-\frac{dL(t)}{dt}$, and $C(t)$=$\frac{1}{\omega_0^2L(t)}$, $\frac{dP(t)}{dt}$ being the polarization current (as $L(t)C(t)$ and $\frac{R(t)}{L(t)}$ remain constant, so do the resonance and collision frequencies). Consequently, the lossless plasma in \cite{1139657} can be modeled as a time-varying RL circuit with a resistor that cancels out the time derivative of the inductor's time-dependence, i.e., $R(t)$=$-\frac{dL(t)}{dt}$.

If we go back to our varying Lorentzian oscillator, we have
\begin{subequations}
\begin{gather}
h(t,\tau)=\chi_h(t,\tau)=A(\tau)\chi(t-\tau),\label{eq:2Z8a}\\
c(t,\hat{\tau})=A(t-\hat{\tau})\chi(\hat{\tau}),\label{eq:2Z8b}
\end{gather}\label{eq:2Z8}%
\end{subequations}
whose Fourier domain representations can be found in Appendix A. Inserting Eq.~(\ref{eq:2Z8a}) into Eq.~(\ref{eq:1})---or Eq.~(\ref{eq:2Z8b}) into Eq.~(\ref{eq:3})---, we arrive at
\begin{equation}
y(t)=\chi(t)\underset{t}{*}\Big(A(t)x(t)\Big),\label{eq:2Z9}
\end{equation}
at which point it appears natural to translate the observation time-frame reference from the origin to $\tau$, when the impulse is applied, and define the so-called output delay-spread function $hc(\hat{\tau},\tau)$=$h(\hat{\tau}+\tau,\tau)$ \cite{1088793}, such that 
\begin{equation}
y(t)=\int_{-\infty}^{\infty}hc(t-\tau,\tau)x(\tau)d\tau,\label{eq:2Z10}
\end{equation}
where $hc(\hat{\tau},\tau)$=$A(\tau)\chi(\hat{\tau})$ is now separable.

\subsection{Polarization in Time-Varying Media}
In \cite{mirmoosa2020dipole}, Mirmoosa et al. investigated the dipolar polazability in time-varying media, and introduced the notion of temporal complex
polarizability.  Here, we take a different path by adopting from  \cite{1454337} the notion of a time-varying admittance for the time-dependent RLC circuit modelling our Lorentzian if we realize that
\begin{equation}
I(t)=\frac{dP(t)}{dt}=\int_{-\infty}^{\infty}\frac{d\chi(t-\tau)}{dt}\frac{1}{L(\tau)}V(\tau)d\tau,\label{eq:2C2}
\end{equation}
in which case $Y_{RLC}(\hat{\tau},\tau)$=$\frac{d\chi(\hat{\tau})}{d\hat{\tau}}\frac{1}{L(\tau)}$. Also, note that this admittance is defined purely in the time domain as a response function. In the next two sections we will delve into the Fourier domain and rigorously characterize the spectra of the impulse response of a time-varying system; this will allow us to utilize generalized time-varying transformed impedances and admittances. But before this, let us first gain physical insights into this problem and consider the simplified case of a time-varying inductor, parameterized by
\begin{equation}
L(t)\frac{dI(t)}{dt}+\frac{dL(t)}{dt}I(t)=E(t),\label{eq:2C3}
%L(t)\frac{d^2P(t)}{dt^2}+\frac{dL(t)}{dt}\frac{dP(t)}{dt}=\epsilon_0E(t)
\end{equation}
where $I(t)$=$\frac{dP(t)}{dt}$. By replacing the right-hand side of the equation above by $\delta(t-\tau)$, the impulse response to this first-order differential equation can be retrieved, for which we first solve the homogeneous equation, which gives us
\begin{equation}
I(t,\tau)=K(\tau)e^{-\int_{0}^{t}\frac{\frac{dL(t')}{dt'}}{L(t')}dt'}=K(\tau)e^{\text{ln}\left(\frac{L(0)}{L(t)}\right)}=K(\tau)\frac{L(0)}{L(t)},\label{eq:2C4}
\end{equation}
with $K(\tau)$ some unknown constant (with respect to $t$), to be determined by imposing $I(\tau,\tau)$=$\frac{1}{L(\tau)}$ \cite{Jeruchim:2000:SCS:557528}. It thus follows that
\begin{equation}
K(\tau)=\frac{1}{L(\tau)}e^{\int_{0}^{\tau}\frac{\frac{dL(t')}{dt'}}{L(t')}dt'}=\frac{1}{L(0)},\label{eq:2C5}
\end{equation}
so $I(t,\tau)$ is simply $\frac{1}{L(t)}$. The impulse response of this system can finally be written as $h_I(t,\tau)$=$I(t,\tau)U(t-\tau)$=$\frac{U(t-\tau)}{L(t)}$ \cite{1454337}. If we assume, e.g., $L(t)$=$\frac{L_0}{1+\Delta\text{cos}(\Omega t)}$, then $K(\tau)$=$\frac{1+\Delta}{L_0}$ and $I(t,\tau)$=$\frac{1+\Delta\text{cos}(\Omega t)}{L_0}$ (note that we could have more easily solved this problem by starting from $\frac{d\Phi(t)}{dt}$=$E(t)$, where $\Phi(t)$=$L(t)I(t)$ represents the magnetic flux linkage, and write $h_I(t,\tau)$=$\frac{h_{\Phi}(t,\tau)}{L(t)}$, with $h_{\Phi}(t,\tau)$=$U(t-\tau)$). By integrating $I(t,\tau)$ with respect to $t$ and enforcing the initial condition of null polarization charge at $t$=$\tau$, we obtain
\begin{equation}
h_{P}(t,\tau)=U(t-\tau)\int_{\tau}^{t}I(t',\tau)dt'=\frac{A(t,\tau)}{L_0}U(t-\tau),\label{eq:2C6}
\end{equation}
with $A(t,\tau)$=$(t-\tau)+\frac{\Delta}{\Omega}\big(\text{sin}(\Omega t)-\text{sin}(\Omega \tau)\big)$. Eq.~(\ref{eq:2C6}) becomes $h_{P}(t,\tau)$=$h_{P}(t-\tau)$=$\frac{t-\tau}{L_0}U(t-\tau)$ when $\Delta$=$0$, with the $t$ term connected to the pole at $\omega$=$0$. It is revealing to compare this expression with the response for the lossless plasma of \cite{1139657}, for which $h_{I}(t,\tau)$=$\frac{U(t-\tau)}{L(\tau)}$ and $h_{P}(t,\tau)$=$\frac{t-\tau}{L(\tau)}U(t-\tau)$, with $L(\tau)$=$\frac{1}{\epsilon_0\omega_p^2(\tau)}$.

If we keep $L(t)$=$\frac{L_0}{1+\Delta\text{cos}(\Omega t)}$ and add a constant resistor $R$, we will obtain
\begin{subequations}
\begin{gather}
h_{I}(t,\tau)=\frac{U(t-\tau)}{L(t)}e^{-\frac{R}{L_0}A(t,\tau)}\label{eq:2C7a}\\
h_{P}(t,\tau)=\frac{U(t-\tau)}{R}\Big(1-e^{-\frac{R}{L_0}A(t,\tau)}\Big).\label{eq:2C7b}
\end{gather}\label{eq:2C7}
\end{subequations}

\comment{
\begin{equation}
\begin{split}
h_{P}(t,\tau)=\frac{1}{R}\Bigg(1-e^{-R L_0\Big((t-\tau)+\frac{\Delta}{\Omega}\big(\text{sin}(\Omega t)-\text{sin}(\Omega \tau)\big)\Big)}\Bigg) \\ \cdot U(t-\tau),\label{eq:2C7}
\end{split}
\end{equation}
}

Similar derivations for an RC circuit with $C(t)$=$\frac{C_0}{1+\Delta\text{cos}(\Omega t)}$ allows us to arrive at
\begin{subequations}
\begin{gather}
h_{P}(t,\tau)=\frac{U(t-\tau)}{R}e^{-\frac{1}{RC_0}A(t,\tau)}\label{eq:2C8a}\\
%h_{I}(t,\tau)=\frac{dh_{P}(t,\tau)}{dt}=\left(\frac{\delta(t-\tau)}{R}-\frac{U(t-\tau)}{R^2C(t)}\right)e^{-\frac{1}{RC_0}A(t,\tau)}.\label{eq:2C8b}
h_{I}(t,\tau)=\frac{dh_{P}(t,\tau)}{dt}=-\frac{U(t-\tau)}{R^2C(t)}e^{-\frac{1}{RC_0}A(t,\tau)}.\label{eq:2C8b}
\end{gather}\label{eq:2C8}
\end{subequations}

\comment{
$c_{I}(t,\hat{\tau})$=$\frac{dh_{P}(t,\tau)}{dt}\Bigr\vert_{\tau=t-\hat{\tau}}$
}

Incidentally, from Eq.~(\ref{eq:3}) one can see that $c_{I}(t,\hat{\tau})\neq\frac{dc_{P}(t,\hat{\tau})}{dt}$, but $c_{I}(t,\hat{\tau})$=$h_{I}(t,t-\hat{\tau})$, just as $hc_{I}(\hat{\tau},\tau)$=$h_{I}(\hat{\tau}+\tau,\tau)$. Obviously, Eq.~(\ref{eq:2C8a}) collapses to Eq.~(\ref{eq:AB4a}) in Appendix B for nondispersive media when $R$=$0$. In addition, in the same way that $\frac{dL(t)}{dt}$ behaves as a resistance, we can observe from the equation below, dual of Eq.~(\ref{eq:2C3}):
\begin{equation}
C(t)\frac{dV(t)}{dt}+\frac{dC(t)}{dt}V(t)=I(t),\label{eq:2C9}
\end{equation}
that $\frac{dC(t)}{dt}$ behaves as a conductance. 

Let us now take a look at the dynamics of an RLC circuit with constant $R$ and $L$, and a capacitor with the same temporal profile as for the previous RC circuit, $C(t)$=$\frac{C_0}{1+\Delta\text{cos}(\Omega t)}$. Repeating the rationale that links Eqs.~(\ref{eq:2Z2}) and (\ref{eq:2Z7}), it is clear that this circuit models the behavior of a medium with a polarization response obeying a Lorentzian curve with varying resonance frequency, described by:
\begin{equation}
\frac{d^2P(t)}{dt^2}+\gamma\frac{dP(t)}{dt}+\omega_0(t)^2P(t)=\epsilon_0\omega_p^2E(t),\label{eq:2C10}
\end{equation}
with $\epsilon_0\omega_p^2$=$\frac{1}{L}$, $\gamma$=$\frac{R}{L}$, and $\omega_0(t)$=$\frac{1}{\sqrt{LC(t)}}$=$\omega_p\sqrt{\frac{\epsilon_0}{C(t)}}$. The homogeneous differential equation for the polarization presents a closed-form solution in terms of even ($M_C$) and odd ($M_S$) Mathieu functions \cite{arfken2013mathematical}, as shown below:
\comment{
\begin{equation}
P(t,\tau)=e^{-\frac{R}{2L}t}\bigg(K_C(\tau)M_C(a,q,z(t))+K_S(\tau)M_S\big(a,q,z(t)\big)\bigg),\label{eq:2C10}
%P(t,\tau)=e^{-\frac{R}{2L}t}[(K_C(\tau)M_C(a,q,z(t))+K_S(\tau)M_S\big(a,q,z(t)\big)],\label{eq:2C10}
\end{equation}
}
\begin{equation}
\begin{split}
P(t,\tau)=&\bigg(K_C(\tau)M_C(a,q,z(t))+K_S(\tau)M_S\big(a,q,z(t)\big)\bigg) \\
&\cdot e^{-\frac{R}{2L}t},\label{eq:2C11}
\end{split}
\end{equation}
with characteristic value $a$=$4\frac{\omega_0^2-\left(\frac{R}{2L}\right)^2}{\Omega^2}$, parameter $q$=$-2\Delta\left(\frac{\omega_0 }{\Omega}\right)^2$, and argument $z(t)$=$\frac{\Omega}{2}t$, as given by the Mathieu differential equation $y''(z)+(a-2q\text{cos}(2z))y(z)$=$0$, $\omega_0$ being $\frac{1}{\sqrt{LC_0}}$. If we enforce $P(t,\tau)$=$0$ and $\frac{dP(t,\tau)}{dt}\Bigr\vert_{t=\tau}$=$\frac{1}{L}$, $K_C(\tau)$ is found to be 
\begin{equation}
K_C(\tau)=\frac{2}{L\Omega}e^{\frac{R}{2L}\tau}\frac{1}{\frac{dM_C(z(t))}{dt}\Bigr\vert_{t=\tau}-\frac{M_C(z(\tau))}{M_S(z(\tau))}\frac{dM_S(z(t))}{dt}\Bigr\vert_{t=\tau}},\label{eq:2C12}
\end{equation}
while $K_S(\tau)$=$-\frac{M_C(z(\tau))}{M_S(z(\tau))}K_C(\tau)$, where the terms $a$ and $q$ have been dropped to simplify the notation. For $\tau$=$0$, $K_C(\tau)$=$0$ and $K_S(\tau)$=$\frac{2}{L\Omega}\frac{1}{\frac{dM_S(z(t))}{dt}\big\vert_{t=0}}$. The time-varying impulse response will finally be $h_P(t,\tau)$=$P(t,\tau)U(t-\tau)$, as was previously done for the varying inductor.

\subsection{Time-Varying Transfer Functions}
We saw before how the polarization/current responses of Eq.~(\ref{eq:2Z2}), or of its RLC circuit equivalent in Eq.~(\ref{eq:2Z7}), do not depend on the medium's state after the impulse and, consequently, derived a separable admittance $Y_{RLC}(\hat{\tau},\tau)$=$\frac{d\chi(\hat{\tau})}{d\hat{\tau}}\frac{1}{L(\tau)}$ which embodies a time-independent frequency dependence. In order to understand what this statement really means, it would be useful to properly define a suitable time-varying transfer function (frequency response). Before going any further, it is expedient to revisit the context of LTV communication channels, whose underlying physical effects are mainly multipath propagation and the Doppler effect, which can be intuitively characterized in terms of time delays and Doppler frequency shifts \cite{hlawatsch2011wireless} (Doppler spectral compression/dilation can be approximated as a frequency shift in narrowband communications), respectively. 

\subsubsection{Transfer Functions for $c(t,\hat{\tau})$}
Denoting by $[\omega,\nu,\hat{\nu}]$ the frequency-domain counterparts of $[t,\tau,\hat{\tau}]$, the specular single-path propagation via an ideal point scatterer $n$ can be captured, except for a complex attenuation constant factor $a_n$, by $c(t,\hat{\tau})$=$e^{i\omega_nt}\delta(\hat{\tau}\!\!-\!\!\hat{\tau}_n)$, which leads to $C_{\omega}(\omega,\hat{\tau})$=$\underset{t\to \omega}{\mathcal{F}\mathcal{T}}\{c(t,\hat{\tau})\}$=$2\pi\delta(\omega\!-\!\omega_n)\delta(\hat{\tau}\!-\!\hat{\tau}_n)$, $\omega_n$ and $\hat{\tau}_n$ being a frequency shift and a time delay, respectively, with $\underset{t\to \omega}{\mathcal{F}\mathcal{T}}$ the Fourier transform (FT) for the ($t$,$\omega$) pair. Following Eq.~(\ref{eq:3}) and using Fubini's theorem \cite{fumni1907sugli}, we can now write 
\begin{equation}
y(t)=\frac{1}{2\pi}\int_{-\infty}^{\infty}\left(\int_{-\infty}^{\infty}C_{\omega}(\omega,\hat{\tau})x(t-\hat{\tau})d\hat{\tau}\right)e^{it\omega}d\omega.\label{eq:2AA1}
\end{equation}

\comment{
\begin{equation}
y(t)=\int_{-\infty}^{\infty}\left(\int_{-\infty}^{\infty}c_{\hat{\tau}}(\hat{\tau})x(t-\hat{\tau})d\hat{\tau}\right)C_{\omega}(\omega)e^{it\omega}d\omega.\label{eq:2AA1}
\end{equation}

\begin{equation}
y(t)=c_{t}(t)\int_{-\infty}^{\infty}\big(C_{\hat{\nu}}(\hat{\nu})X(\hat{\nu})\big)e^{it\hat{\nu}}d\hat{\nu}.\label{eq:2AA1}
\end{equation}
}

That is, the integral along $\omega$ can be viewed as a continuous parallel connection of LTI channels, each parameterized by a Doppler frequency $\omega$ (note that the term inside the parentheses depends on $t$, so $\int_{-\infty}^{\infty}\left(\right)e^{it\omega}d\omega$ in Eq.~(\ref{eq:2AA1}) is not an inverse FT). Analogously, if we flip domains on both dimensions and define $C_{\hat{\nu}}(t,\hat{\nu})$=$\underset{\hat{\tau}\to\hat{\nu}}{\mathcal{F}\mathcal{T}}\{c(t,\hat{\tau})\}$=$e^{i\omega_nt}e^{-i\hat{\tau}\hat{\nu}}$, after manipulating Eq.~(\ref{eq:3}) it can be shown that
\begin{equation}
\begin{split}
y(t)&=\frac{1}{2\pi}\int_{-\infty}^{\infty}C_{\hat{\nu}}(t,\hat{\nu})\left(\int_{-\infty}^{\infty}x(t-\hat{\tau})e^{i\hat{\tau}\hat{\nu}}d\hat{\tau}\right)d\hat{\nu} \\
&=\frac{1}{2\pi}\int_{-\infty}^{\infty}\big(C_{\hat{\nu}}(t,\hat{\nu})X(\hat{\nu})\big)e^{it\hat{\nu}}d\hat{\nu},\label{eq:2AA2}
\end{split}
\end{equation}
where $X(\omega)$=$\underset{t\to \omega}{\mathcal{F}\mathcal{T}}\{x(t)\}$ and, again and for the same reason, $\int_{-\infty}^{\infty}\left(\right)e^{it\hat{\nu}}d\hat{\nu}$ is not an inverse FT. Finally, using both transformed domains and $C_{\omega,\hat{\nu}}(\omega,\hat{\nu})$=$\underset{(t,\hat{\tau})\to(\omega,\hat{\nu})}{\mathcal{F}\mathcal{T}}\{c(t,\hat{\tau})\}$=$2\pi\delta(\omega\!\!\!\!-\!\!\!\!\omega_n)e^{-i\hat{\tau}\hat{\nu}}$, it is easy to arrive at
\begin{equation}
Y(\omega)=\frac{1}{2\pi}\int_{-\infty}^{\infty}C_{\omega,\hat{\nu}}(\omega-\hat{\nu},\hat{\nu})X(\hat{\nu})d\hat{\nu}.\label{eq:2AA3}
\end{equation}

$C_{\omega}(\omega,\hat{\tau})$ describes how the input signal is spread out or broadened both in frequency ($\omega$) and time ($\hat{\tau}$), whereas $C_{\hat{\nu}}(t,\hat{\nu})$ expresses the response's time ($t$) and frequency ($\hat{\nu}$) selectivity. For an LTI system, there is no $\omega$-broadening or $t$-selectivity, so $C_{\omega}(\omega,\hat{\tau})$ and $C_{\hat{\nu}}(t,\hat{\nu})$ are simplified to $\delta(\omega)c(\hat{\tau})$ and $C_{\hat{\nu}}(\hat{\nu})$, respectively. Note also that, if $c(t,\hat{\tau})$ is separable,  $C_{\omega,\hat{\nu}}(\omega,\hat{\nu})$=$C_{\omega}(\omega)C_{\hat{\nu}}(\hat{\nu})$  and thus Eq.~(\ref{eq:2AA3}) can be simplified as
\begin{equation}
Y(\omega)=\frac{1}{2\pi}C_{\omega}(\omega)\underset{\omega}{*}\big(C_{\hat{\nu}}(\omega)X(\omega)\big),\label{eq:2AA4}
\end{equation}
which is the frequency-domain version of Eq.~(\ref{eq:5}).

\subsubsection{Transfer Functions for $hc(\hat{\tau},\tau)$}
Although a detailed description of the transfer functions of $h(t,\tau)$ and $hc(\hat{\tau},\tau)$ can be found in Appendices A and B, respectively, it is worthy to focus on $hc(\hat{\tau},\tau)$ and see that Eqs.~(\ref{eq:2AA3})---and (\ref{eq:AA2}) in Appendix A---can be rewritten as
\begin{equation}
Y(\omega)=\frac{1}{2\pi}\int_{-\infty}^{\infty}HC_{\hat{\nu},\nu}(\omega,\nu+\omega)X(-\nu)d\nu,\label{eq:2AC3}
\end{equation}
which, if $hc(\hat{\tau},\tau)$ is separable, i.e., $hc(\hat{\tau},\tau)$=$hc_{\hat{\tau}}(\hat{\tau})hc_{\tau}(\tau)$ and thus $HC_{\hat{\nu},\nu}(\hat{\nu},\nu)$=$HC_{\hat{\nu}}(\hat{\nu})HC_{\nu}(\nu)$, adopts the form
\begin{equation}
Y(\omega)=\frac{1}{2\pi}HC_{\hat{\nu}}(\omega)\left(HC_{\nu}(\omega)\underset{\omega}{*}X(\omega)\right),\label{eq:2AC4}
\end{equation}
which is the FT of Eq.~(\ref{eq:2Z9}) if we note that $\chi(\hat{\tau})$=$hc_{\hat{\tau}}(\hat{\tau})$ and $A(\tau)$=$hc_{\tau}(\tau)$. This
shows an interesting duality between the pairs of Eqs.~(\ref{eq:2Z9},\ref{eq:2AC4}) and (\ref{eq:5},\ref{eq:2AA4}).

Continuing with our varying Lorentzian, we have
\begin{equation}
HC_{\nu}(\hat{\tau},\nu)=A(\nu)\chi(\hat{\tau}),\;\;\;\;HC_{\hat{\nu}}(\hat{\nu},\tau)=A(\tau)\chi(\hat{\nu}),\label{eq:2AD1}
\end{equation}
where we see the convenience of working with the ($\hat{\tau}$,$\tau$) pair (note that, although $A$ in $A(\nu)$ is the FT of $A$ in $A(\tau)$, we deliberately choose to not add more notation and let its argument resolve the ambiguity. The same applies to $\chi$, and to the circuital elements $R$, $L$, and $C$ in the next section). This stems from the fact that frequency broadening (time selectivity), in sheer contrast with the Doppler $\omega$-spreading ($t$-selectivity) defined so far, is now given in the $\nu$ ($\tau$) domain.
The time-variance of the Doppler channel entails $\omega$-broadening, whereas the Lorentzian's varying nature reveals itself in the width of $N(\nu)$. Note also that if one replaces $HC_{\nu}$=$A(\nu)$ and $HC_{\hat{\nu}}(\hat{\nu})$=$\chi(\hat{\nu})$=$\frac{1}{\omega_0^2-\hat{\nu}^2+i\gamma\hat{\nu}}$ in Eq.~(\ref{eq:2AC4}), what is obtained is precisely Eq.~(\ref{eq:2Z3}), except for the constant $\epsilon_0$.

\subsection{Time-varying Impedances and Admittances}
Now that we have discussed a mathematical theory of LTV systems, we can utilize, as in \cite{1454337}, the notion of time-varying impedance for our RLC circuit's time-varying impedance. It is clear that $Z_R(t,\hat{\tau})$=$R(t)\delta(\hat{\tau})$, $Z_C(t,\hat{\tau})$=$\frac{U(\hat{\tau})}{C(t)}$ and
\begin{equation}
Z_L(t,\hat{\tau})=L(t-\hat{\tau})\delta'(\hat{\tau})=L(t)\delta'(\hat{\tau})+\frac{dL(t)}{dt}\delta(\hat{\tau}),\label{eq:2B1}
\end{equation}
where, incidentally, note that the delta function and all of its derivatives are causal distributions \cite{beerends2003fourier}. Therefore, we can write the transformed impedances as
\begin{subequations}
\begin{gather}
Z_{R\omega}(\omega,\hat{\tau})=R(\omega)\delta(\hat{\tau}),\label{eq:2B2a} \\
Z_{L\omega}(\omega,\hat{\tau})=L(\omega)\big(\delta'(\hat{\tau})+i\omega\delta(\hat{\tau})\big),\label{eq:2B2b} \\
Z_{C\omega}(\omega,\hat{\tau})=\underset{t\to \omega}{\mathcal{F}\mathcal{T}}\left\{\frac{1}{C(t)}\right\}U(\hat{\tau}),\label{eq:2B2c}
\end{gather}\label{eq:2B2}%
\end{subequations}
and
\begin{subequations}
\begin{gather}
Z_{R\hat{\nu}}(t,\hat{\nu})=R(t),\label{eq:2B3a} \\
Z_{L\hat{\nu}}(t,\hat{\nu})=L(t)i\hat{\nu}+\frac{dL(t)}{dt},\label{eq:2B3b} \\
Z_{C\hat{\nu}}(t,\hat{\nu})=\frac{1}{C(t)}\left(\frac{1}{i\hat{\nu}}+\pi\delta(\hat{\nu})\right),\label{eq:2B3c}
\end{gather}\label{eq:2B3}%
\end{subequations}
or in the ($\omega$,$\hat{\nu}$)-domain as
\begin{subequations}
\begin{gather}
Z_{R\omega,\hat{\nu}}(\omega,\hat{\nu})=R(\omega),\label{eq:2B4a} \\
Z_{L\omega,\hat{\nu}}(\omega,\hat{\nu})=iL(\omega)\big(\hat{\nu}+\omega \big),\label{eq:2B4b} \\
Z_{C\omega,\hat{\nu}}(\omega,\hat{\nu})=\underset{t\to\omega}{\mathcal{F}\mathcal{T}}\left\{\frac{1}{C(t)}\right\}\left(\frac{1}{i\hat{\nu}}+\pi\delta(\hat{\nu})\right).\label{eq:2B4c}
\end{gather}\label{eq:2B4}
\end{subequations}

These expressions clearly show how $\omega$-dispersion ($\hat{\nu}$-dispersion) in $Z_{L\omega}$ ($Z_{L\hat{\nu}}$) depends on $\hat{\tau}$ ($t$). By duality, the same interdependence shows up in the capacitor's admittance (see Appendix C). It is paramount to realize, though, that the usual time-invariant relation between impedance and admittance does not apply now and, consequently, it cannot be used to circumvent the lack of closed-form solution of, e.g., the FT of Eqs.~(\ref{eq:2C7},\ref{eq:2C8}). For instance, noting that $h_{I}(t,\hat{\tau})$ in Eq.~(\ref{eq:2C8b}) is the time-varying admittance response of the RC circuit, and switching to the ($t$,$\hat{\tau}$)-space, we have
\begin{subequations}
\begin{gather}
Y_{RC\omega}(\omega,\hat{\tau})\neq \big(Z_{R\omega}(\omega,\hat{\tau})+Z_{C\omega}(\omega,\hat{\tau})\big)^{-1},\label{eq:2BB1a} \\
Y_{RC\hat{\nu}}(t,\hat{\nu})\neq \big(Z_{R\hat{\nu}}(t,\hat{\nu})+Z_{C\hat{\nu}}(t,\hat{\nu})\big)^{-1},\label{eq:2BB1b} \\
Y_{RC\omega,\hat{\nu}}(\omega,\hat{\nu})\neq\big(Z_{R\omega,\hat{\nu}}(\omega,\hat{\nu})+Z_{C\omega,\hat{\nu}}(\omega,\hat{\nu})\big)^{-1}.\label{eq:2BB1c}
\end{gather}\label{eq:2BB1}%
\end{subequations}
where, e.g., $Y_{RC\omega}(\omega,\hat{\tau})$=$\underset{t\to \omega}{\mathcal{F}\mathcal{T}}\{h_{I}(t,\hat{\tau})\}$.

Nonetheless, the impedance of the series RLC circuit that models the Lorentzian resulting from a time-varying $N(t)$ can be written in separable form, considering that $R(\omega)=(\gamma-i\omega)L(\omega)$ and $\underset{t\to \omega}{\mathcal{F}\mathcal{T}}\left\{\frac{1}{C(t)}\right\}=\omega_0^2L(\omega)$, as:
\begin{subequations}
\begin{gather}
Z_{RLC}(t,\hat{\tau})=L(t)Z_{RLC\hat{\tau}}(\hat{\tau}),\label{eq:2B5a} \\
Z_{RLC\omega,\hat{\nu}}(\omega,\hat{\nu})=L(\omega)Z_{RLC\hat{\nu}}(\hat{\nu}),\label{eq:2B5b}
\end{gather}\label{eq:2B5}%
\end{subequations}
where $L(t)$ plays the role of $Z_{RLCt}(t)$ (the definition of $Z_{RLC\omega}(\omega,\hat{\tau})$ and $Z_{RLC\hat{\nu}}(t,\hat{\nu})$ is straightforward and thus omitted for brevity), and with
\begin{subequations}
\begin{gather}
Z_{RLC\hat{\tau}}(\hat{\tau})=\gamma\delta(\hat{\tau})+\delta'(\hat{\tau})+\omega_0^2U(\hat{\tau}),\label{eq:2B6a} \\
Z_{RLC\hat{\nu}}(\hat{\nu})=\gamma+i\hat{\nu}+\omega_0^2\left(\frac{1}{i\hat{\nu}}+\pi\delta(\hat{\nu})\right),\label{eq:2B6b}
\end{gather}\label{eq:2B6}%
\end{subequations}
being the impedance of a time-invariant RLC circuit with normalized elements $R\!=\!\gamma\!=\!\frac{1}{L(t)}\left(R(t)+\frac{dL(t)}{dt}\right)$, $L\!\!=\!\!1$, and $C\!\!=\!\!\frac{1}{\omega_0^2}$. The impedance's $\hat{\nu}$-dispersion is $t$-independent (the whole $\hat{\nu}$-spectrum is modulated by the same factor $L(t)$), just like $\hat{\tau}$-broadening is $\omega$-independent. Besides, inserting  Eq.~(\ref{eq:2B5b}) in Eq.~(\ref{eq:2AA3}), it is easy to see that
\begin{equation}
V(\omega)=\frac{1}{2\pi}L(\omega)\underset{\omega}{*}\big(Z_{RLC\hat{\nu}}(\omega)I(\omega)\big)\label{eq:2B7}
\end{equation}
and therefore
\begin{equation}
V(t)=L(t)\Big(Z_{RLC\hat{\tau}}(t)\underset{t}{*}I(t)\Big),\label{eq:2B8}
\end{equation}
both consistent with Eqs.~(\ref{eq:2AA4}) and (\ref{eq:5}), respectively.

That is, the voltage response at the observation instant $t$ is the product of $L(t)$ times the convolution of the input current with the response of an LTI, in this case a ``normalized'' RLC circuit. Except for the terms $\gamma$ and $i\omega$ (or, in the time domain, $\gamma\delta(t)$ and $\delta'(t)$) which represent, respectively, the instantaneous response of the time-invariant normalized resistor $R\!=\!\gamma$ and inductor $L\!=\!1$, $Z_{RLC\hat{\tau}}(t)\underset{t}{*}i(t)$ is simply the ratio of the total charge accumulated in the capacitor and $C\!=\!\frac{1}{\omega_0^2}$, i.e., its voltage. We previously showed how the current response of our varying Lorentzian at $t$ to a voltage impulse at $\tau$ is only a function of the system's state at $\tau$; now we observe the opposite behavior: the voltage response at $t$ to a current impulse at $\tau$ is only a function of the system's state at $t$. This interrelation is best seen by reordering and Fourier-transforming Eq.~(\ref{eq:2B8}) to arrive at
\begin{equation}
I(\omega)=\frac{1}{2\pi}Y_{RLC\hat{\nu}}(\omega)\left(\underset{t\to\omega}{\mathcal{F}\mathcal{T}}\left\{\frac{1}{L(t)}\right\}\underset{\omega}{*}V(\omega)\right),\label{eq:2BB2}
\end{equation}
where we have used the LTI equality $Y_{RLC\hat{\nu}}(\hat{\nu})\!=\!Z_{RLC\hat{\nu}}^{-1}(\hat{\nu})$, which does hold now. Going back to the time domain, we have
\begin{equation}
I(t)=Y_{RLC\hat{\tau}}(t)\underset{t}{*}\left(\frac{1}{L(t)}V(t)\right).\label{eq:2BB3}
\end{equation}
This last pair of equations has precisely the form of Eqs.~(\ref{eq:2Z9},\ref{eq:2AC4}), as expected. Finally, we can write
\begin{subequations}
\begin{gather}
Y_{RLC}(\hat{\tau},\tau)=\frac{1}{L(\tau)}Y_{RLC\hat{\tau}}(\hat{\tau}),\label{eq:2BB4a} \\
Y_{RLC\hat{\nu},\nu}(\hat{\nu},\nu)=\underset{\tau\to\nu}{\mathcal{F}\mathcal{T}}\left\{\frac{1}{L(\tau)}\right\}Y_{RLC\hat{\nu}}(\hat{\nu}),\label{eq:2BB4b}
\end{gather}\label{eq:2B45}%
\end{subequations}
and realize that $Y_{RLC\hat{\tau}}(\hat{\tau})$=$\frac{d\chi(\hat{\tau})}{d\hat{\tau}}$ when the integration constant $P_0$ in $P(t)$=$P_0+\int_{-\infty}^{t}I(\tau)d\tau$, which translates into the term $\pi\delta(\hat{\nu})$ within $Z_{RLC\hat{\nu}}(\hat{\nu})$, is omitted. The admittance's $\hat{\nu}$-dispersion is $\tau$-independent (the entire $\hat{\nu}$-spectrum is now modulated by the same factor $L(\tau)$).
\comment{
where we have dropped the $\pi\delta(\omega)$ term of $\underset{\hat{\tau}\to \hat{\nu}} {\mathcal{F}\mathcal{T}}\{U(\hat{\tau})\}$, assuming there is no DC component.
}
\comment{
\begin{equation}
\frac{\partial ^2}{\partial t^2}\left(\frac{\partial ^2P(z,t)}{\partial z^2}-\frac{1}{c^2}\frac{\partial ^2P(z,t)}{\partial t^2}\right)=\frac{1}{c^2}N(t)\frac{\partial ^2P(z,t)}{\partial t^2},\label{eq:2BB6}
\end{equation}
}

\subsection{Kramers-Kronig Relations}
Given that the Kramers-Kronig relations \cite{deL.Kronig:26,kramers1927diffusion} connect the real and imaginary parts of any complex function that is analytic in the upper half-plane, and that, for any stable physical system, causality implies analyticity and vice versa, we now explore these relations when the system is time-varying. In the context of time-invariant media, it is well known that the Kramers-Kronig relations constitute a powerful tool to retrieve the real part of the permittivity from absorption measurements (e.g., electron energy loss spectroscopy \cite{PhysRevB.30.1155}). Moreover, they prove useful in obtaining the real part of the effective nonlinear change of permittivity from its imaginary part (e.g., in the case of metals, via the change of interband transitions involving Fermi-level states \cite{Guo2016}), in which case the medium's nonlinear response is slow enough to consider it effectively time-invariant.

Going back to our time-varying impulse responses, it was pointed out before that a causal LTV system requires $h(t,\tau)$ to be zero for $t<\tau$, and thereby $c(t,\hat{\tau})$ (and $hc(\hat{\tau},\tau)$) must also be zero for $\hat{\tau}<0$. Ergo, it is evident that the Kramers-Kronig relations have physical ground along the $\hat{\nu}$-dimension. In the ($t$,$\hat{\nu}$)-space and considering $c(t,\hat{\tau})$ first, we will have $t$-varying Kramers-Kronig relations of the form
\begin{subequations}
\begin{gather}
\text{Re}\left\{C_{\hat{\nu}}(t,\hat{\nu})\right\}=\frac{1}{\pi}\dashint_{-\infty}^{\infty}\frac{\text{Im}\left\{C_{\hat{\nu}}(t,\hat{\nu}')\right\}}{\hat{\nu}-\hat{\nu}'}d\hat{\nu}',\label{eq:301a} \\
\text{Im}\left\{C_{\hat{\nu}}(t,\hat{\nu})\right\}=-\frac{1}{\pi}\dashint_{-\infty}^{\infty}\frac{\text{Re}\left\{C_{\hat{\nu}}(t,\hat{\nu}')\right\}}{\hat{\nu}-\hat{\nu}'}d\hat{\nu}',\label{eq:301b}
\end{gather}\label{eq:301}%
\end{subequations}
or, alternatively, in the ($\omega$,$\hat{\nu}$)-space, $\omega$-dependent Kramers-Kronig relations as shown below
\begin{subequations}
\begin{gather}
\text{Re}\left\{C_{\omega,\hat{\nu}}(\omega,\hat{\nu})\right\}=\frac{1}{\pi}\dashint_{-\infty}^{\infty}\frac{\text{Im}\left\{C_{\omega,\hat{\nu}}(\omega,\hat{\nu}')\right\}}{\hat{\nu}-\hat{\nu}'}d\hat{\nu}',\label{eq:302a} \\
\text{Im}\left\{C_{\omega,\hat{\nu}}(\omega,\hat{\nu})\right\}=-\frac{1}{\pi}\dashint_{-\infty}^{\infty}\frac{\text{Re}\left\{C_{\omega,\hat{\nu}}(\omega,\hat{\nu}')\right\}}{\hat{\nu}-\hat{\nu}'}d\hat{\nu}',\label{eq:302b}
\end{gather}\label{eq:302}%
\end{subequations}
where $\dashint$ stands for the Cauchy principal value of the integral. In this regard, note that $C_{\omega}(\omega,\hat{\tau})$ is not purely real in general, but this fact does not compromise the validity of Eq.~(\ref{eq:302}). For each value of $\omega$, one could use superposition and apply the Hilbert transform to the real and imaginary parts of the spectra of $\text{Re}\left\{C_{\omega}(\omega,\hat{\tau})\right\}$ and $i\text{Im}\left\{C_{\omega}(\omega,\hat{\tau})\right\}$ separately, the only difference in the latter being the purely imaginary/real character of the $\hat{\nu}$-spectra of its even/odd decomposition. The drawback of investigating the causality of $c(t,\hat{\tau})$ along the $\hat{\tau}$-axis is that, in actuality, we are analyzing the response of the system (medium) at a fixed $t$ when considering all possible delays, which does not give an intuition of the system's dynamics to a single impulse response (the explanation of Fig.~\ref{fig:2} in the next section reveals this fact in more detail). It is therefore more suitable in our case to resort to $hc(\hat{\tau},\tau)$ and write (in the following, only the retrieval of the real part from the imaginary part is included for brevity):
\begin{subequations}
\begin{gather}
\text{Re}\left\{HC_{\hat{\nu}}(\hat{\nu},\tau)\right\}=\frac{1}{\pi}\dashint_{-\infty}^{\infty}\frac{\text{Im}\left\{HC_{\hat{\nu}}(\hat{\nu}',\tau)\right\}}{\hat{\nu}-\hat{\nu}'}d\hat{\nu}',\label{eq:303a} \\
\text{Re}\left\{HC_{\hat{\nu},\nu}(\hat{\nu},\nu)\right\}=\frac{1}{\pi}\dashint_{-\infty}^{\infty}\frac{\text{Im}\left\{HC_{\hat{\nu},\nu}(\hat{\nu}',\nu)\right\}}{\hat{\nu}-\hat{\nu}'}d\hat{\nu}'.\label{eq:303b}
\end{gather}\label{eq:303}
\end{subequations}
Using Eqs.~(\ref{eq:AB1a}) and Eq.~(\ref{eq:AB1c}) in Appendix B one can still derive the following: 
\begin{subequations}
\begin{gather}
\text{Re}\left\{e^{i\tau\omega}H_{\omega}(\omega,\tau)\right\}=\frac{1}{\pi}\dashint_{-\infty}^{\infty}\frac{\text{Im}\left\{e^{i\tau\omega'}H_{\omega}(\omega',\tau)\right\}}{\omega-\omega'}d\omega',\label{eq:304a} \\
\text{Re}\left\{H_{\omega,\nu}(\omega,\nu-\omega)\right\}=\frac{1}{\pi}\dashint_{-\infty}^{\infty}\frac{\text{Im}\left\{H_{\omega,\nu}(\omega',\nu-\omega')\right\}}{\omega-\omega'}d\omega',\label{eq:304b}
\end{gather}\label{eq:304}%
\end{subequations}
where Eq.~(\ref{eq:304a}) can also be obtained by simply decomposing $h(t,\tau)$ into even and odd with respect to $t$=$\tau$. A similar expression can be derived for $H_{\nu}(t,\nu)$ using anticausality (the expressions that relate the real and imaginary parts of an anticausal---not to be confused with noncausal---signal's spectrum are the same as for a causal signal, but with the signs flipped) and symmetry with respect to $\tau$=$t$.

\comment{
\begin{subequations}
\begin{gather}
\text{Re}\left\{HC_{\hat{\nu}}(\hat{\nu},\tau)\right\}=\frac{1}{\pi}\dashint_{-\infty}^{\infty}\frac{\text{Im}\left\{HC_{\hat{\nu}}(\hat{\nu}',\tau)\right\}}{\hat{\nu}'-\hat{\nu}}d\hat{\nu}',\label{eq:303a} \\
\text{Im}\left\{HC_{\hat{\nu}}(\hat{\nu},\tau)\right\}=-\frac{1}{\pi}\dashint_{-\infty}^{\infty}\frac{\text{Re}\left\{HC_{\hat{\nu}}(\hat{\nu}',\tau)\right\}}{\hat{\nu}'-\hat{\nu}}d\hat{\nu}',\label{eq:303b}
\end{gather}\label{eq:303}
\end{subequations}

\begin{subequations}
\begin{gather}
\text{Re}\left\{HC_{\hat{\nu},\nu}(\hat{\nu},\nu)\right\}=\frac{1}{\pi}\dashint_{-\infty}^{\infty}\frac{\text{Im}\left\{HC_{\hat{\nu},\nu}(\hat{\nu}',\nu)\right\}}{\hat{\nu}'-\hat{\nu}}d\hat{\nu}',\label{eq:304a} \\
\text{Im}\left\{HC_{\hat{\nu},\nu}(\hat{\nu},\nu)\right\}=-\frac{1}{\pi}\dashint_{-\infty}^{\infty}\frac{\text{Re}\left\{HC_{\hat{\nu},\nu}(\hat{\nu}',\nu)\right\}}{\hat{\nu}'-\hat{\nu}}d\hat{\nu}',\label{eq:304b}
\end{gather}\label{eq:304}
\end{subequations}

\begin{subequations}
\begin{gather}
\text{Re}\left\{e^{i\tau\omega}H_{\omega}(\omega,\tau)\right\}=\frac{1}{\pi}\dashint_{-\infty}^{\infty}\frac{\text{Im}\left\{e^{i\tau\omega'}H_{\omega}(\omega',\tau)\right\}}{\omega'-\omega}d\omega',\label{eq:305a} \\
\text{Im}\left\{e^{i\tau\omega}H_{\omega}(\omega,\tau)\right\}=-\frac{1}{\pi}\dashint_{-\infty}^{\infty}\frac{\text{Re}\left\{e^{i\tau\omega'}H_{\omega}(\omega',\tau)\right\}}{\omega'-\omega}d\omega',\label{eq:305b}
\end{gather}\label{eq:305}
\end{subequations}

\begin{subequations}
\begin{gather}
\text{Re}\left\{H_{\omega,\nu}(\omega,\nu-\omega)\right\}=\frac{1}{\pi}\dashint_{-\infty}^{\infty}\frac{\text{Im}\left\{H_{\omega,\nu}(\omega',\nu-\omega')\right\}}{\omega'-\omega}d\omega',\label{eq:306a} \\
\text{Im}\left\{H_{\omega,\nu}(\omega,\nu-\omega)\right\}=-\frac{1}{\pi}\dashint_{-\infty}^{\infty}\frac{\text{Re}\left\{H_{\omega,\nu}(\omega',\nu-\omega')\right\}}{\omega'-\omega}d\omega',\label{eq:306b}
\end{gather}\label{eq:306}
\end{subequations}
}

\section{Numerical Results}

\begin{figure*}[h!]
\includegraphics[width=7in]{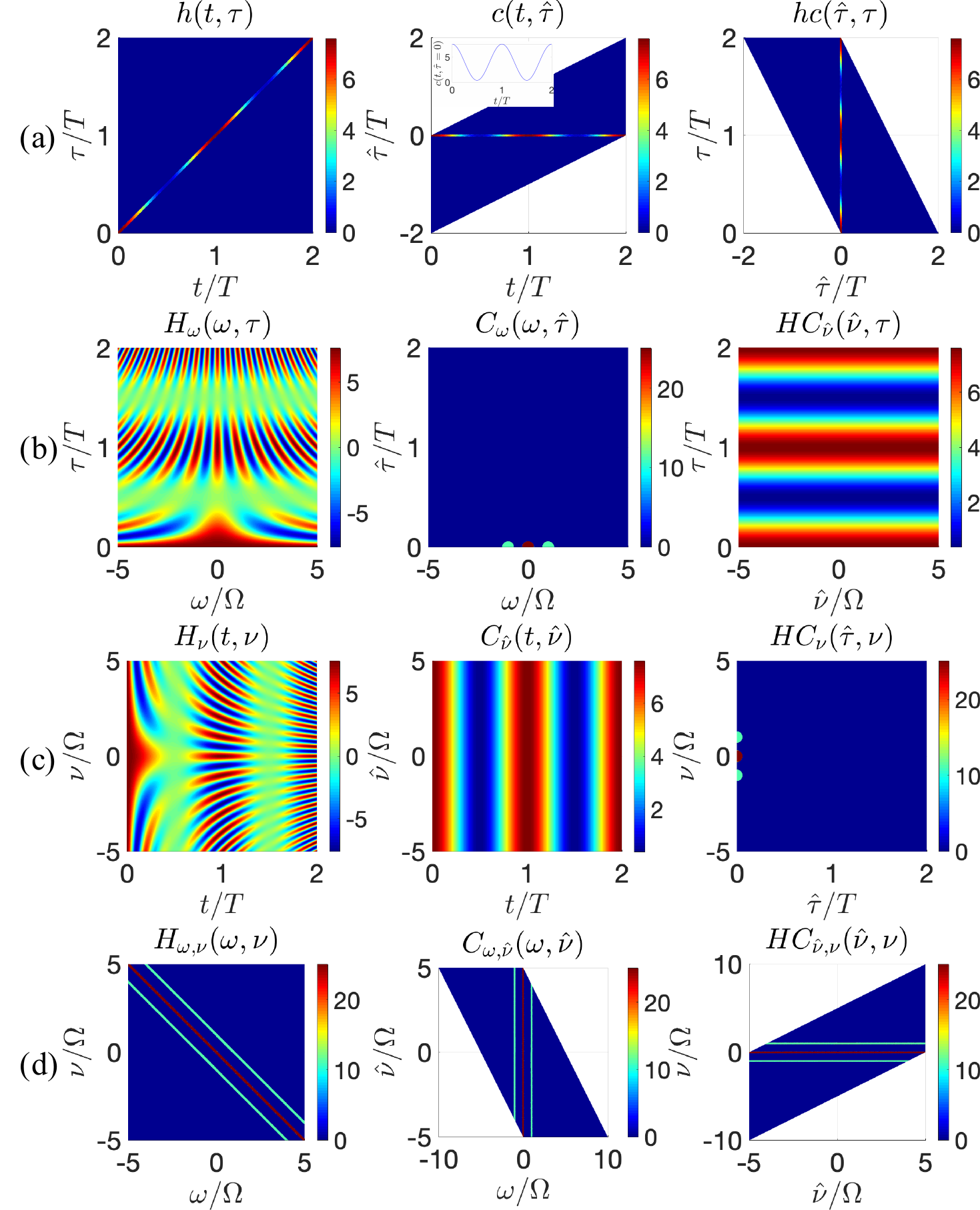}% Here is how to import EPS art
%\captionsetup{width=2\linewidth}
%\captionsetup{width=7in}
\caption{Time-varying impulse responses $h$, $c$ and $hc$ for a nondispersive medium of polarization described by $P(t)$=$\epsilon_0C(t)E(t)$. (a) Both the observation time $t$-axis and the impulse time $\tau$-axis are represented in the temporal domain. $c(t,\hat{\tau})$ and $hc(\hat{\tau},\tau)$ are remapped from $h(t,\tau)$ to better illustrate the effect of changing the representation spaces. The inset in $c(t,\hat{\tau})$ represents $c(t,\hat{\tau}=0)$=$C(t)$ (b) The $t$-axis is transformed to $\omega$, while the $\tau$-axis is left unchanged. (c) The $t$-axis stays in the time-domain, while the $\tau$-axis is transformed to $\nu$. (d) Both time axes are transformed. $C_{\omega,\hat{\nu}}$ and $HC_{\hat{\nu},\nu}$ are remapped from $H_{\omega,\nu}$. Only the real part of the spectra is depicted in (b-d). In (b,c) the magnified dots represent Dirac delta functions.}
\label{fig:1}
\end{figure*}

In order to visualize the relations between $h$, $c$ and $hc$, both in time and frequency, in Fig.~\ref{fig:1} we first consider the trivial scenario of the nondispersive time-varying medium of Eqs.~(\ref{eq:AB4},\ref{eq:AB5}) and choose, for simplicity, $C(t)$=$C_0\big(1+\Delta\text{cos}(\Omega t)\big)$, such that $C(\omega)$=$C_0\pi\Big(2\delta(\omega)+\Delta\big(\delta(\omega-\Omega)+\delta(\omega+\Omega)\big)\Big)$, with $C_0$=$4$ and $\Delta$=$0.9$. It is evident that $c$ or $hc$ are much more convenient than $h$, specially in (b,c). 

% [h,t,b,p]: [here, top, bottom, page of float]

\begin{figure*}[h]
\includegraphics[width=7in]{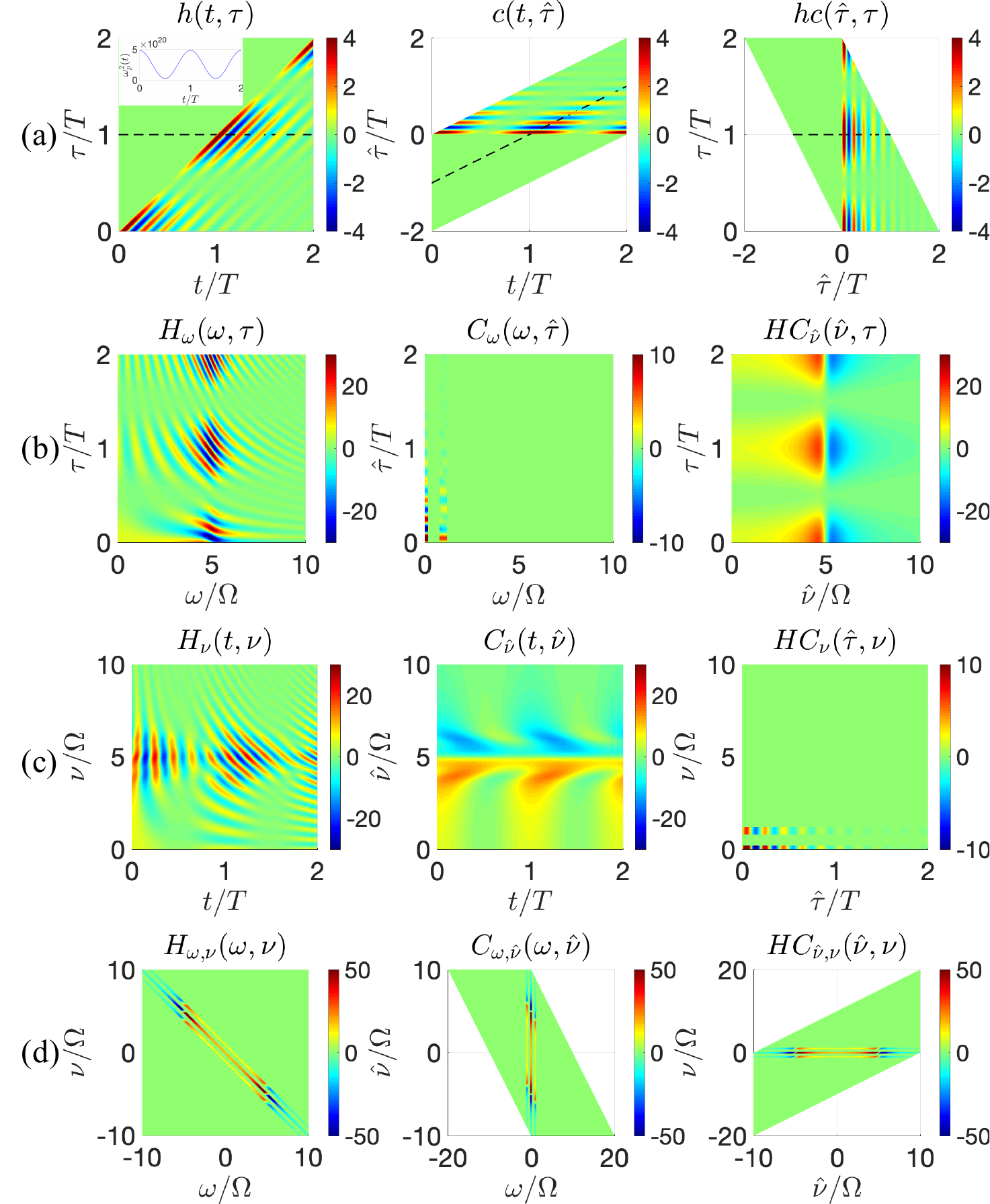}% Here is how to import EPS art
\caption{Time-varying impulse responses $h$, $c$ and $hc$ for a medium with time-varying Lorentzian response, with $\tau$-independent $\hat{\nu}$-dispersion. Panels (a-d) are organized in the same way as Fig.~\ref{fig:1}. The inset in (a) for $h(t,\tau)$ represents the varying plasma frequency $\omega_p^2(t)$ in [rad/s]. The magnitudes of the response functions where $\chi$ shows up in the time domain are divided by $\omega_0$ (similar considerations apply to Fig.~\ref{fig:4} and Fig.~\ref{fig:5}).}
\label{fig:2}
\end{figure*}

As a second example, let us now consider in Fig.~\ref{fig:2} a time-varying medium whose electric polarization follows Eqs.~(\ref{eq:2Z2},\ref{eq:2Z3}), and focus on its Lorentzian-like impulse responses as defined in Eqs.~(\ref{eq:2Z8},\ref{eq:2AD1}), with a $\hat{\nu}$-dispersion that remains unchanged regardless of $\tau$. The plasma frequency is chosen to be periodically modulated as $\omega_p^2(t)$=$\omega_{p0}^2\big(1+\text{cos}(\Omega t)\big)$, with $\omega_{p0}$ (and $\gamma$) such that, in the LTI case of $\Delta$=$0$, $\chi(\Omega)$=$3-0.10i$, with $\omega_0$=$5\Omega$. $\Delta$ has the same value as in Fig.~\ref{fig:1}.

The black dashed straight lines in panel (a) of Fig.~\ref{fig:2} (and Fig.~\ref{fig:4}) illustrate how points in the $(t,\tau)$-domain are remapped onto the $(t,\hat{\tau})$ and $(\hat{\tau},\tau)$ domains. If we have an input impulse at a given $\tau_0$, the information about the causal's system's response can be found in $h(t\!>\!\tau_0,\tau_0)$, which is a straight line parallel to the $t$-axis. This same information can also be found in $c(t,t\!-\!\tau_0\!>\!0)$, which forms a straight line at an angle of $45^{\circ}$ with respect to the $t$-axis, crossing it at $t$=$\tau_0$. One-dimensional (1D) cuts of $c(t,\hat{\tau})$ parallel to the $t$-axis restrict the response of the system vs. $t$ to only a given delay $\hat{\tau}$, implicitly implying an input to the system that is a continuous train of impulses at $t-\hat{\tau}$. This is very clearly visualized, except for the cosinusoidal variation, in panel (a) of Fig.~\ref{fig:1} for the case of a medium with instantaneous response. Analogously, 1D cuts of $c(t,\hat{\tau})$ parallel to the $\hat{\tau}$-axis describe the response of the system at a given $t$ for all possible delays, again implying a constant input.

On the contrary, the system's response for an impulse at $\tau$=$\tau_0$ is also contained in $hc(t\!-\!\tau_0\!>\!0,\tau_0)$, thereby drawing a straight line parallel to the $t$-axis, just as with $h(t,\tau)$, but with the advantage that now the causality condition is $\tau$-independent. 1D cuts of $hc(\hat{\tau},\tau)$ parallel to the $\hat{\tau}$-axis have a less useful meaning, as they characterize the system response for a given delay $\hat{\tau}$, which entails the aforementioned constant input. Incidentally, note also that $hc(\hat{\tau},\tau)$ in panel (a) is $\tau$-periodic.

\begin{figure*}[h]
\includegraphics[width=7in]{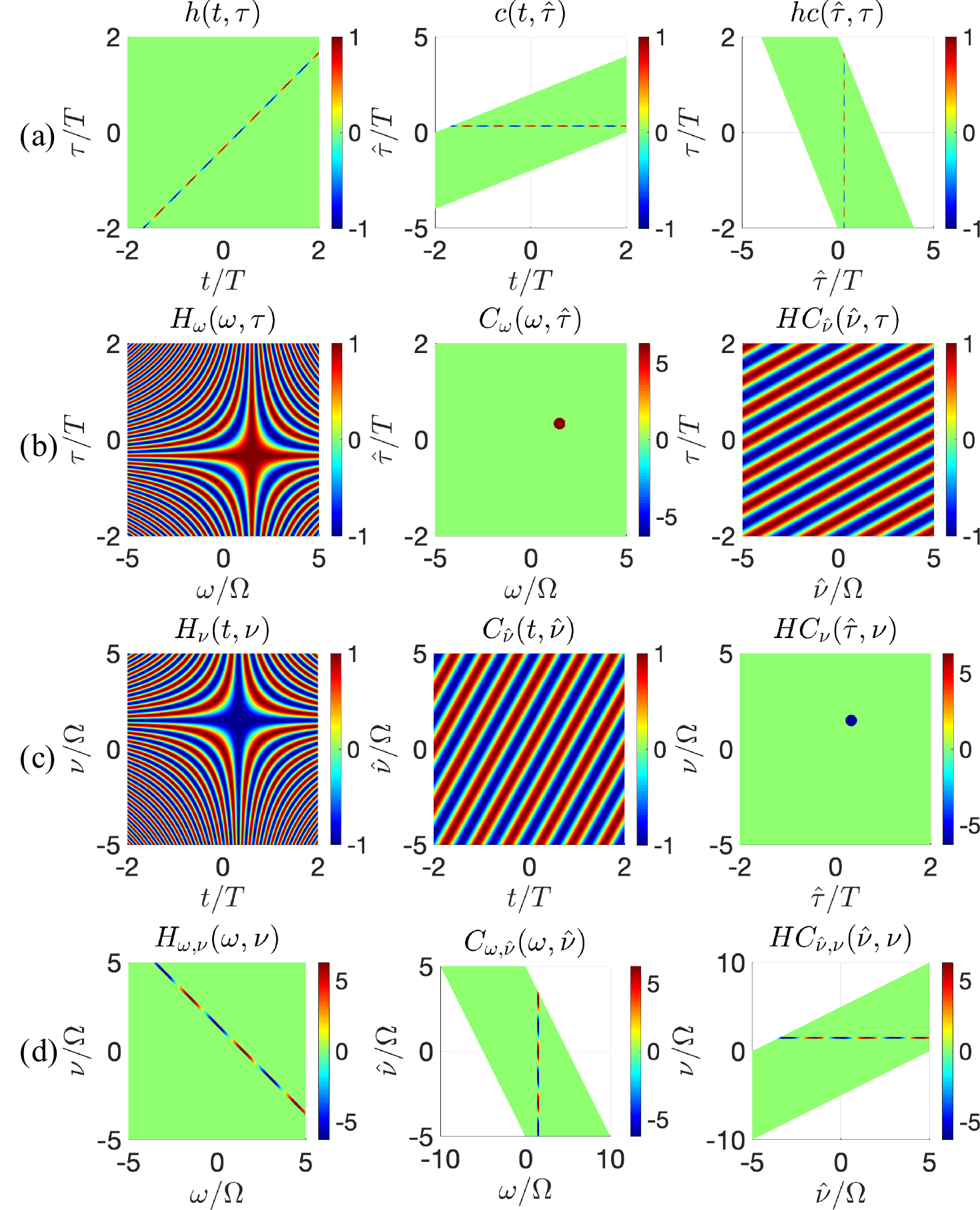}% Here is how to import EPS art
\caption{Time-varying impulse responses $h$, $c$ and $hc$ for a single-path propagation channel affected by Doppler shift. Panels (a-d) are organized in the same way as the previous figures.}
\label{fig:3}
\end{figure*}

\begin{figure*}[h]
\includegraphics[width=7in]{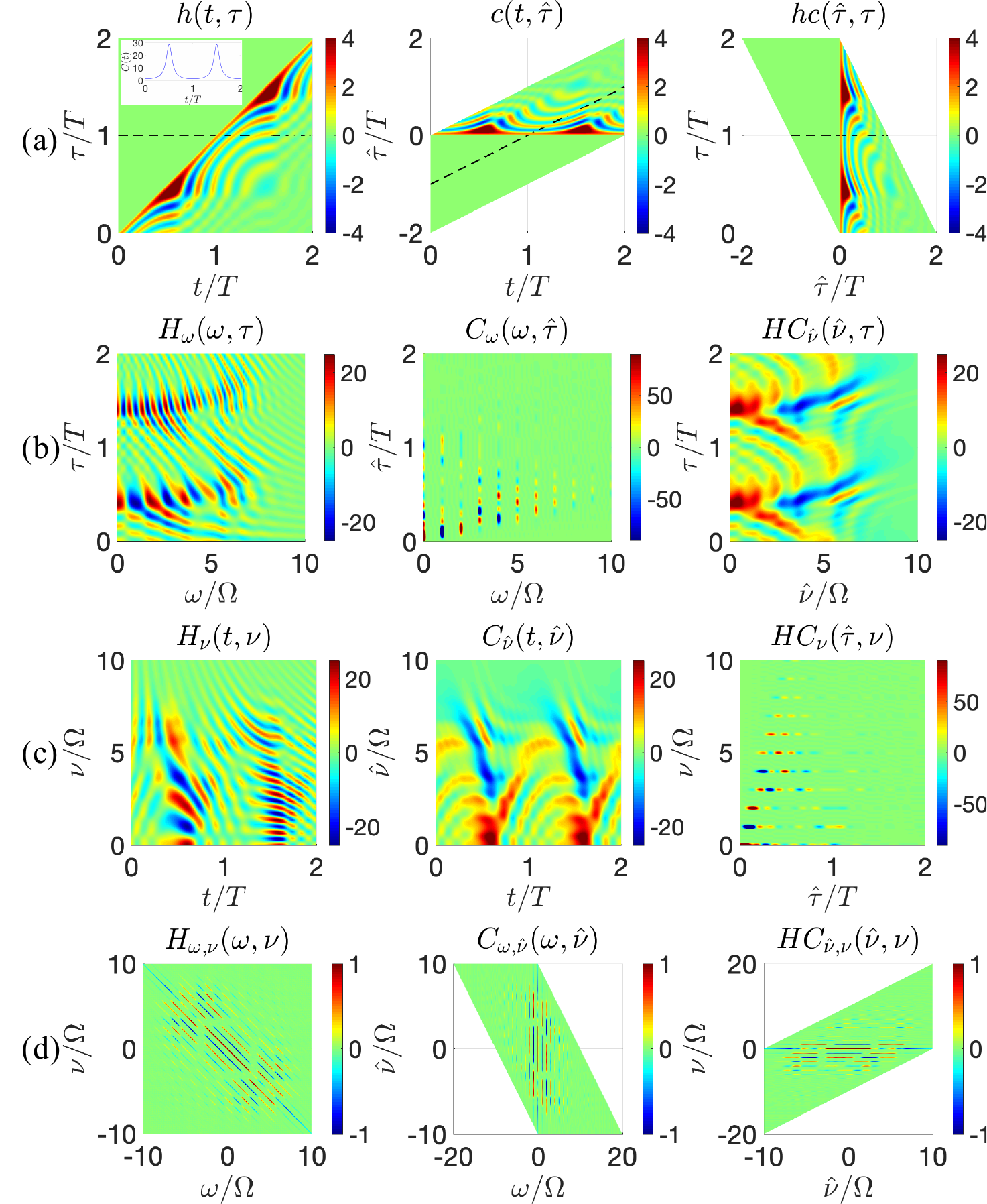}% Here is how to import EPS art
\caption{Panels (a-d) display, in the same order as before, the impulse responses $h$, $c$ and $hc$ for the time-variant medium with polarization charge characterized by Eq.~(\ref{eq:2B6}). Unlike Figs.~\ref{fig:1}- $\!\!$\ref{fig:3}, which are entirely analytical, the plots in the spectral domains are now calculated numerically through fast Fourier transforms.}
\label{fig:4}
\end{figure*}

In short, $c(t,\hat{\tau})$ is a powerful tool in mobile communications because $c(t=t_0,\hat{\tau})$ synthesizes, at a given instant $t_0$, the signal at the receiver including all the delays; whereas $hc(\hat{\tau},\tau)$ is more revealing in our case because $hc(\hat{\tau},\tau=\tau_0)$ tells us what the response of the varying medium is for a single impulse occurring at $\tau_0$. We already addressed the advantage of the former for mobile communications, when using the simplified case $c(t,\hat{\tau})$=$e^{i\omega_nt}\delta(\hat{\tau}-\hat{\tau}_n)$ (see Eq.~(\ref{eq:2AA1}), e.g.), which we now depict in Fig.~\ref{fig:3} for $\omega_n$=$1.5\Omega$ and $\hat{\tau}_n$=$\frac{T}{3}$, with $T$=$\frac{2\pi}{\Omega}$. One can observe that our first example in Fig.~\ref{fig:1} can actually be recast into the shape of a zero-delay ($\hat{\tau}_n$=0) three-path channel with Doppler shifts $0$, $+\Omega$ and $-\Omega$.

\begin{figure*}[h]
\includegraphics[width=7in]{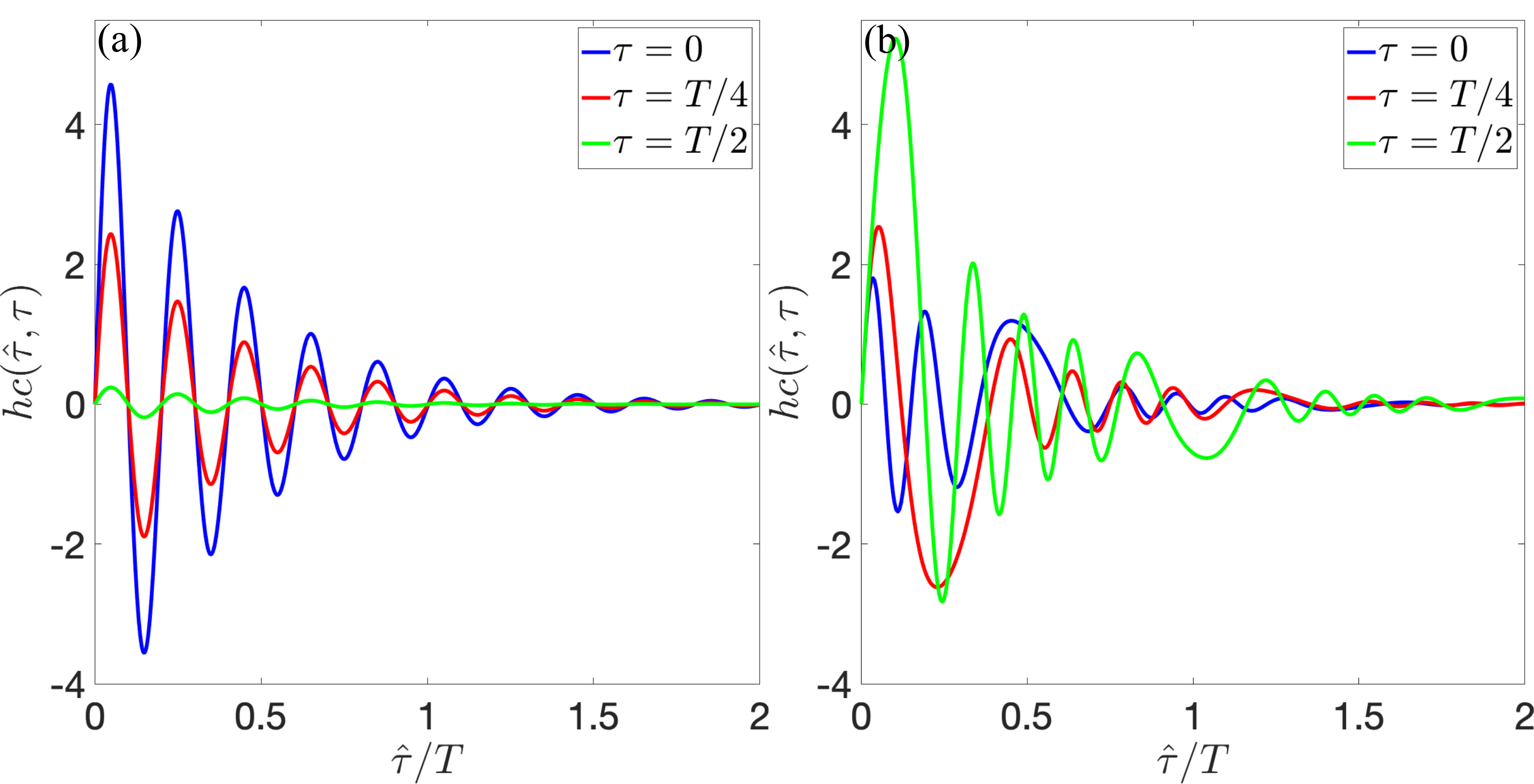}% Here is how to import EPS art
\caption{1D cuts of $hc(\hat{\tau},\tau)$ for different fixed values of $\tau$. Panel (a) and (b) correspond to Figs.~\ref{fig:2} and \ref{fig:4}, respectively. Panel (a) shows how the system response is the same regardless of $\tau$, except for a constant. Meanwhile, panel (b) presents three curves with totally different shape, revealing how $\hat{\nu}$-dispersion does depend on $\tau$ in this case.}
\label{fig:5}
\end{figure*}

Fig.~\ref{fig:4} represents the medium whose polarization response is Lorentzian with time-varying resonance frequency, according to Eq.~(\ref{eq:2B6}). As mentioned earlier, this is equivalent to an RLC circuit with a time-dependent capacitor. In a similar fashion as Fig.~\ref{fig:2}, the capacitor is periodically modulated as $C(t)$=$\frac{C_0}{1+\Delta\text{cos}(\Omega t)}$, with $C_0$ (and $\gamma$) such that, in the LTI case of $\Delta$=$0$, $\chi(\Omega)$=$3-0.10i$, with $\omega_0$=$5\Omega$. We choose again $\Delta$=$0.9$. Note that with this modulation, $\omega_0^2(t)$ follows a sinusoidal pattern. Panel (b) in Fig.~\ref{fig:5} illustrates how the dispersion of this medium varies with $\tau$, unlike the medium with varying plasma frequency of Fig.~\ref{fig:2}, represented in panel (a) of Fig.~\ref{fig:5}. Importantly, the response to an impulse applied at $\tau$ is now a function of how $C(t)$ evolves for $t>\tau$.

\comment{
It would probably be advantageous in some cases to work with the one-sided Laplace transform with respect to $\hat{\tau}$, and with the FT with respect to either $t$ or $\tau$.
}

\section{Conclusions}

In this work we have borrowed the mathematical framework that rigorously characterizes LTV systems in the signal processing research field, of particular interest in mobile communication channels, and adapted/extended it to address the topic of time-variant generally-dispersive electromagnetic constitutive responses. In doing so we have shown that the concept of time-varying frequency dispersion is still physically meaningful when the medium's temporal variation is fast with respect to the driving field's frequency. 
In LTI systems, it is very well known that the causality of its impulse response allows to relate the real and imaginary part of its spectra through the Hilbert transform or, in the jargon of the physics community, the Kramers-Kronig relations. We herein described the response of a causal LTV system as a differential equation with time-varying coefficients and linked these coefficients to time-dependent lumped circuital elements. We then defined the different Fourier-transformed spaces that result from a twofold temporal variation: for each observation instant $\tau$, the system has a different impulse response, each of them having an LTI equivalence when expressed with respect to $t-\tau$. We proved that these Fourier spaces give room to time-varying transfer functions for which not only is it possible to generalize the Kramers-Kronig relations, but also allow us to utilize the generalized impedance and admittance of varying resistors, inductors and capacitors. Furthermore, as an example of medium with time-varying dielectric response, we studied the Lorentzian dispersion resulting from a varying number of polarizable atoms $N(t)$; interestingly, we saw that the dielectric response of such medium to an impulse applied at $\tau$ is only a function of $N(t\!=\!\tau)$ and not of $N(t\!>\!\tau)$.

\begin{acknowledgments}
This work is supported in part by the Vannevar Bush Faculty Fellowship program sponsored by the Basic Research Office of the Assistant Secretary of Defense for Research and Engineering and funded by the Office of Naval Research through grant N00014-16-1-2029. 
\end{acknowledgments}

\appendix                                     
\section{Transfer Functions for $h(t,\tau)$}
The impulse response $h(t,\tau)$ can be transformed under $\underset{t\to \omega}{\mathcal{F}\mathcal{T}}$, $\underset{\tau\to\nu}{\mathcal{F}\mathcal{T}}$ and $\underset{(t,\tau)\to(\omega,\nu)}{\mathcal{F}\mathcal{T}}$ to yield $H_{\omega}(\omega,\tau)$, $H_{\nu}(t,\nu)$ and $H_{\omega,\nu}(\omega,\nu)$, respectively, and express
\begin{equation}
\begin{split}
y(t)&=\frac{1}{2\pi}\int_{-\infty}^{\infty}\left(\int_{-\infty}^{\infty}H_{\omega}(\omega,\tau)x(\tau)d\tau\right)e^{it\omega}d\omega \\
&=\frac{1}{2\pi}\int_{-\infty}^{\infty}H_{\nu}(t,\nu)X(-\nu)d\nu\label{eq:AA1}
\end{split}
\end{equation}
and
\begin{equation}
Y(\omega)=\frac{1}{2\pi}\int_{-\infty}^{\infty}H_{\omega,\nu}(\omega,\nu)X(-\nu)d\nu.\label{eq:AA2}
\end{equation}

Likewise, it is straightforward to infer that
\begin{subequations}
\begin{gather}
H_{\nu}(t,\nu)=e^{-it\nu}C_{\hat{\nu}}(t,-\nu),\label{eq:AA3a}\\
H_{\omega,\nu}(\omega,\nu)=C_{\omega,\hat{\nu}}(\omega+\nu,-\nu).\label{eq:AA3b}
\end{gather}\label{eq:AA3}
\end{subequations}
On the contrary, the link in the ($\omega$,$\tau$) domain is more intricate and reads $H_{\omega}(\omega,\tau)$=$\frac{1}{2\pi}e^{-i\tau\omega}\underset{\omega,\tau}{*}C_{\omega}(\omega,-\tau)$, $\underset{\omega,\tau}{*}$ denoting a double convolution operation across the $\omega$ and $\tau$ dimensions. This convoluted connection can be traced back to the fact that frequency $\omega$-broadening loses its meaning when switching from $\hat{\tau}$ to $\tau$. This is better illustrated if we take a look at $h(t,\tau)$=$c(t,t\!-\!\tau)$=$e^{i\omega_nt}\delta(t\!-\!\tau\!-\!\hat{\tau}_n)$, with $t$ now showing up both within the complex exponential and the Dirac delta function, rendering $H_{\omega}(\omega,\tau)$=$e^{-i(\omega-\omega_n)(\tau+\hat{\tau}_n)}$. Going back to our varying Lorentzian oscillator, from Eqs.~(\ref{eq:2Z8}) it follows that
\begin{subequations}
\begin{gather}
H_{\omega}(\omega,\tau)=A(\tau)\chi(\omega)e^{-i\tau\omega},\label{eq:AA5a}\\
C_{\omega}(\omega,\hat{\tau})=A(\omega)\chi(\hat{\tau})e^{-i\hat{\tau}\omega}.\label{eq:AA5b}
\end{gather}\label{eq:AA5}
\end{subequations}
As far as the complementary domains are concerned, we now have
\begin{subequations}
\begin{gather}
H_{\nu}(t,\nu)=\frac{1}{2\pi}A(\nu)\underset{\nu}{*}\left(\chi(-\nu)e^{-it\nu}\right),\label{eq:AA6a}\\
C_{\hat{\nu}}(t,\hat{\nu})=\frac{1}{2\pi}\left(A(-\hat{\nu})e^{-it\hat{\nu}}\right)\underset{\hat{\nu}}{*}\chi(\hat{\nu}),\label{eq:AA6b}
\end{gather}\label{eq:AA6}%
\end{subequations}
where it becomes apparent that Eq.~(\ref{eq:AA3a}) is satisfied if one realizes that
\begin{equation}
\begin{split}
C_{\hat{\nu}}(t,\hat{\nu})&=\frac{1}{2\pi}\int_{-\infty}^{\infty}A(-\hat{\nu}-\hat{\nu}')e^{-it(\hat{\nu}-\hat{\nu}')}\chi(\hat{\nu}')d\hat{\nu}' \\
&=\frac{1}{2\pi}e^{-it\hat{\nu}}\int_{-\infty}^{\infty}A(-\hat{\nu}-\hat{\nu}')e^{it\hat{\nu}'}\chi(\hat{\nu}')d\hat{\nu}' \\
&=\frac{1}{2\pi}e^{-it\hat{\nu}}\left(A(-\hat{\nu})\underset{\hat{\nu}}{*}\left(\chi(\hat{\nu})e^{it\hat{\nu}}\right)\right).\label{eq:AA7}
\end{split}
\end{equation}
In addition, one can also write
\begin{subequations}
\begin{gather}
H_{\omega,\nu}(\omega,\nu)=A(\nu+\omega)\chi(\omega),\label{eq:AA8a}\\
C_{\omega,\hat{\nu}}(\omega,\hat{\nu})=A(\omega)\chi(\hat{\nu}+\omega).\label{eq:AA8b}
\end{gather}\label{eq:AA8}
\end{subequations}

\comment{
\begin{equation}
y(t)=\int_{-\infty}^{\infty}\left(\int_{-\infty}^{\infty}HC_{\hat{\nu}}(\omega,\tau)e^{-i\tau\omega}x(\tau)d\tau\right)e^{it\omega}d\omega,\label{eq:AA9}
\end{equation}
}

\section{Transfer Functions for $hc(\hat{\tau},\tau)$}
From $hc(\hat{\tau},\tau)$=$h(\hat{\tau}+\tau,\tau)$, we can find that 
\begin{subequations}
\begin{gather}
HC_{\hat{\nu}}(\hat{\nu},\tau)=e^{i\tau\hat{\nu}}H_{\omega}(\hat{\nu},\tau),\label{eq:AB1a}\\
HC_{\nu}(\hat{\tau},\nu)=\frac{1}{2\pi}e^{i\hat{\tau}\nu}\underset{\hat{\tau},\nu}{*}H_{\nu}(\hat{\tau},\nu),\label{eq:AB1b}\\
HC_{\hat{\nu},\nu}(\hat{\nu},\nu)=H_{\omega,\nu}(\hat{\nu},\nu-\hat{\nu}).\label{eq:AB1c}
\end{gather}\label{eq:AB1}%
\end{subequations}
\balance
Equivalently, $hc(\hat{\tau},\tau)$=$c(\hat{\tau}+\tau,\hat{\tau})$, and hence
\begin{subequations}
\begin{gather}
HC_{\hat{\nu}}(\hat{\nu},\tau)=\frac{1}{2\pi}e^{i\tau\hat{\nu}}\underset{\hat{\nu},\tau}{*}C_{\hat{\nu}}(\tau,\hat{\nu}),\label{eq:AB2a}\\
HC_{\nu}(\hat{\tau},\nu)=e^{i\hat{\tau}\nu}C_{\omega}(\nu,\hat{\tau}),\label{eq:AB2b}\\
HC_{\hat{\nu},\nu}(\hat{\nu},\nu)=C_{\omega,\hat{\nu}}(\nu,\hat{\nu}-\nu),\label{eq:AB2c}
\end{gather}\label{eq:AB2}%
\end{subequations}
which means we can rewrite Eq.~(\ref{eq:2AA1}) as
\begin{equation}
y(t)=\frac{1}{2\pi}\int_{-\infty}^{\infty}\left(\int_{-\infty}^{\infty}e^{-i\hat{\tau}\omega}HC_{\nu}(\hat{\tau},\omega)x(t-\hat{\tau})d\hat{\tau}\right)e^{it\omega}d\omega,\label{eq:AB3}
\end{equation}
that is to say, as a continuous sum of parallel LTI channels, each with an impulse response $h(t)$=$e^{-it\omega}HC_{\nu}(t,\omega)$. What we cannot do is to reformulate the rightmost part of Eq.~(\ref{eq:2AA2}) or Eq.~(\ref{eq:AA1}) in terms of $HC_{\hat{\nu}}$ or $HC_{\nu}$, respectively, because time $\hat{\tau}$-broadening is not physically meaningful anymore, as can be seen in our Doppler LTV channel, where $hc(\hat{\tau},\tau)$=$e^{i\omega_n(\hat{\tau}+\tau)}\delta(\hat{\tau}\!-\!\hat{\tau}_n)$.

If we assume a nondispersive (instantaneous) time-varying medium with a varying capacitor, as given by $P(t)$=$\epsilon_0C(t)E(t)$, it is clear that
\begin{subequations}
\begin{gather}
h(t,\tau)=C(\tau)\delta(t-\tau)=C(t)\delta(t-\tau),\label{eq:AB4a}\\
c(t,\hat{\tau})=C(t-\hat{\tau})\delta(\hat{\tau})=C(t)\delta(\hat{\tau}),\label{eq:AB4b}\\
hc(\hat{\tau},\tau)=C(\tau)\delta(\hat{\tau}),\label{eq:AB4c}
\end{gather}\label{eq:AB4}%
\end{subequations}
so there is not much difference between expressing the system's time-dependence with respect to $t$ or $\tau$, other than temporal shifts. In the transformed domains, we would have
\begin{subequations}
\begin{gather}
H_{\omega}(\omega,\tau)=C(\tau)e^{-i\tau\omega},\;\;\;\;H_{\nu}(t,\nu)=C(t)e^{-it\nu}\label{eq:AB5a}\\
C_{\omega}(\omega,\hat{\tau})=C(\omega)\delta(\hat{\tau}),\;\;\;\;C_{\hat{\nu}}(t,\hat{\nu})=C(t),\label{eq:AB5b}\\
HC_{\hat{\nu}}(\hat{\nu},\tau)=C(\tau),\;\;\;\;HC_{\nu}(\hat{\tau},\nu)=C(\nu)\delta(\hat{\tau}).\label{eq:AB5c}
\end{gather}\label{eq:AB5}%
\end{subequations}

\section{Time-varying Admittances}
Given that $Y_R(\hat{\tau},\tau)$=$\frac{1}{R(\tau)}\delta(\hat{\tau})$, $Y_L(\hat{\tau},\tau)$=$\frac{U(\hat{\tau})}{L(\tau+\hat{\tau})}$ and $Y_C(\hat{\tau},\tau)$=$C(\tau)\delta'(\hat{\tau})$, the transformed admittances become
\comment{
\begin{equation}
Y_C(\hat{\tau},\tau)=C(\tau)\delta'(\hat{\tau})=C(\tau+\hat{\tau})\delta'(\hat{\tau})+C'(\tau)\delta(\hat{\tau}),\label{eq:AC1}
\end{equation}
}
\begin{subequations}
\begin{gather}
Y_{R\hat{\nu}}(\hat{\nu},\tau)=\frac{1}{R(\tau)},\label{eq:AC2a} \\
Y_{L\hat{\nu}}(\hat{\nu},\tau)=\frac{1}{2\pi}\left(\underset{\hat{\tau}\to \hat{\nu}}{\mathcal{F}\mathcal{T}}\left\{\frac{1}{L(\hat{\tau})}\right\}e^{i\tau\hat{\nu}}\right)\underset{\hat{\nu}}{*}\left(\frac{1}{i\hat{\nu}}+\pi\delta(\hat{\nu})\right),\label{eq:AC2b} \\
Y_{C\hat{\nu}}(\hat{\nu},\tau)=C(\tau)i\hat{\nu},\label{eq:AC2c}
\end{gather}\label{eq:AC2}%
\end{subequations}
and
\begin{subequations}
\begin{gather}
Y_{R\nu}(\hat{\tau},\nu)=\underset{\tau\to \nu}{\mathcal{F}\mathcal{T}}\left\{\frac{1}{R(\tau)}\right\}\delta(\hat{\tau}),\label{eq:AC3a} \\
Y_{L\nu}(\hat{\tau},\nu)=\underset{\tau\to \nu}{\mathcal{F}\mathcal{T}}\left\{\frac{1}{L(\tau)}\right\}e^{i\hat{\tau}\nu}U(\hat{\tau}),\label{eq:AC3b} \\
Y_{C\nu}(\hat{\tau},\nu)=C(\nu)\delta'(\hat{\tau}),\label{eq:AC3c}
\end{gather}\label{eq:AC3}%
\end{subequations}
or in the ($\hat{\nu}$,$\nu$)-domain as
\begin{subequations}
\begin{gather}
Y_{R\hat{\nu},\nu}(\hat{\nu},\nu)=\underset{\tau\to \nu}{\mathcal{F}\mathcal{T}}\left\{\frac{1}{R(\tau)}\right\},\label{eq:AC4a} \\
Y_{L\hat{\nu},\nu}(\hat{\nu},\nu)=\underset{\tau\to \nu}{\mathcal{F}\mathcal{T}}\left\{\frac{1}{L(\tau)}\right\}\left(\frac{1}{i(\hat{\nu}-\nu)}+\pi\delta(\hat{\nu}-\nu)\right),\label{eq:AC4b} \\
Y_{C\hat{\nu},\nu}(\hat{\nu},\nu)=C(\nu)i\hat{\nu}.\label{eq:AC4c}
\end{gather}\label{eq:AC4}
\end{subequations}
In addition, the expressions for the admittances in the ($t$,$\hat{\tau}$)-space and its transformed counterparts result from applying duality to Eqs.~(\ref{eq:2B2})-(\ref{eq:2B4}).

%\section{CACA}
%jajaja

\nocite{*}

\bibliography{apssamp}% Produces the bibliography via BibTeX.

\end{document}